\documentclass[aps,prx,twocolumn,longbibliography,superscriptaddress,nofootinbib]{revtex4-2}

\usepackage{graphicx}% Include figure files
\usepackage{dcolumn}% Align table columns on decimal point
\usepackage{bm}% bold math
\usepackage[english]{babel}
\usepackage{lipsum} % Used for inserting dummy 'Lorem ipsum' text into the template
\usepackage{hyperref}
\usepackage{times}
\usepackage{amsfonts,amssymb,bm}
\usepackage{amsmath}
\usepackage{epsfig}
\usepackage{mathrsfs}
\usepackage{color,soul}
\usepackage{bbold}
\hypersetup{colorlinks,%
citecolor=black,%
filecolor=black,%
linkcolor=black,%
urlcolor=black}
\usepackage{pstricks}
\usepackage{multirow}
\usepackage{graphicx}
\usepackage{textcomp}
\usepackage{braket}
\usepackage{empheq}
\usepackage[normalem]{ulem}
\definecolor{color1bg}{HTML}{b890a1}
\usepackage{hyperref}
\hypersetup{
 colorlinks,
 citecolor=color1bg,
 filecolor=black,
 linkcolor=color1bg,
 urlcolor=black}
\usepackage{color}
\usepackage{rotating}
\usepackage{amsmath}
\usepackage{amsthm}
\usepackage{eurosym}
\usepackage{siunitx}
\usepackage{rotating}
\usepackage{amsmath}
\usepackage{eurosym}
\usepackage[babel]{csquotes}

\usepackage[font=small]{caption}
\usepackage{mwe}
\DeclareMathOperator{\tr}{Tr}

\usepackage[framemethod=default]{mdframed}
\usepackage[labelfont=bf,labelsep=quad,justification=raggedright]{caption}
\usepackage{newfloat}

\DeclareFloatingEnvironment[name={S}]{suppfigure}
\DeclareFloatingEnvironment[name={S}]{supptable}
\allowdisplaybreaks
\usepackage{titlecaps}% http://ctan.org/pkg/titlecaps

\newmdenv[linecolor=white,backgroundcolor=gray!15!]{myframe}

\definecolor{green1}{rgb}{0.33, 0.7, 0.69}

\begin{document}

\title{Coarse-grained quantum thermodynamics: \\ Observation-dependent quantities, observation-independent laws}

\author{Giulia Rubino}
\thanks{giulia.rubino@bristol.ac.uk}
\affiliation{Quantum Engineering Technology Labs, H. H. Wills Physics Laboratory and School of Electrical, Electronic, and Mechanical Engineering, University of Bristol, BS8 1FD, UK}
\affiliation{H. H. Wills Physics Laboratory, University of Bristol, Tyndall Avenue, Bristol, BS8 1TL, United Kingdom}

\author{\v{C}aslav Brukner}
\thanks{caslav.brukner@univie.ac.at}
\affiliation{Vienna Center for Quantum Science and Technology (VCQ), Faculty of Physics, University of Vienna, Boltzmanngasse 5, 1090, Vienna, Austria}
\affiliation{Institute for Quantum Optics and Quantum Information (IQOQI), Austrian Academy of Sciences, Boltzmanngasse 3, 1090 Vienna, Austria}

\author{Gonzalo Manzano}
\thanks{gmanzano@ucm.es}
\affiliation{Institute for Cross-Disciplinary Physics and Complex Systems (IFISC) UIB-CSIC, Campus Universitat Illes Balears, E-07122 Palma de Mallorca, Spain}

\date{\today}
\begin{abstract}
In both classical and quantum thermodynamics, physical quantities are typically assigned objective values defined independently of our observations. We then refer to the `work performed by a gas', or the `entropy of the gas', regardless of how they are evaluated. Here, we question this conception in the context of quantum thermodynamics, estimating how the definition of pivotal thermodynamic quantities is affected by experimental instruments of limited precision. We find that the coarse-grained thermodynamic quantities frequently lead to different conclusions from those drawn in fine-grained scenarios. For instance, the irreversibility of a process, or its work payoff, can significantly vary with the instrument precision. We show nonetheless that coarse-grained thermodynamic quantities satisfy the \textit{same} relations  (\textit{i.e.}, the second law inequality, the relation between dissipation and distinguishability of a process from its time-reverse, and the quantum work fluctuation theorems) as their fine-grained counterparts. These results highlight the observation-independence of relations linking thermodynamic quantities which are themselves observation-dependent.
\end{abstract}
\maketitle

In traditional formulations of thermodynamics, notions such as thermal equilibrium, thermodynamic entropy, or extractable work are often introduced as objective macroscopic properties of a system, without explicit reference to the information available to an observer.
In quantum mechanics, by contrast, the notion of quantum states and their dynamics is more explicitly understood as reflecting the information that an agent is able to acquire about the system.
This shift in perspective inspires the question of whether a similar epistemic viewpoint might also be meaningful (or even necessary) in the thermodynamic context. 
In fact, in real experiments, agents do not dispose of instruments of infinite precision. Consequently, physical properties can only be reconstructed in a coarse-grained manner, which inherently limits the amount of information that can be acquired.
This limitation takes on particular importance in thermodynamics, where key quantities like entropy and work are fundamentally tied to information.
Such considerations gave rise to the field of information thermodynamics, which examines how constraints on information processing shape fundamental thermodynamic limits. 
A paradigmatic example is the modern understanding of Maxwell's demon, where the extractable work reflects the level of knowledge the agent has about the system \cite{MaxwellDemon,Parrondo2015,Goold_2016}. From this perspective, there appears to be no inherent meaning in, \textit{e.g.}, the notion of `work performed on or by a gas', or of the `thermodynamic entropy of the gas' \emph{per se}, without specifying the observer's level of knowledge and their experimental ability to engineer the state of the gas.

While a redefinition of the main thermodynamic quantities under coarse-graining emerges as a natural response to these considerations, it raises a number of related questions such as how these quantities evolve with varying experimental precision, what energetic costs are associated with coarse-graining, and whether thermodynamic laws remain invariant under such changes. Providing answers to these questions is the central purpose of this study.

Various different types of coarse-graining procedures have already been explored in the thermodynamic context for small classical and quantum systems, following either Hamiltonian dynamics~\cite{Kawai2007,Parrondo2009,Horowitz:2009} or stochastic evolution~\cite{Rahav:07,GomezMarin:08,Vulpiani:10,Esposito:12,Vollmer:12,Esposito:15}. Most of these approaches focus on eliminating fast degrees of freedom in the dynamics~\cite{Celani:17,Seifert19} or on marginal observations of selected states or currents~\cite{Shiriashi:15,Polettini:17,Bisker:17,Manzano24,Ferri25}. In contrast, here we develop a framework for quantum coarse-graining that explicitly models the finite precision of measurement devices, representing their limited energy resolution through a family of finite-degree projectors that decimate the system’s Hilbert space. This coarse-graining homogenises blocks within the density operator according to the chosen energy resolution, providing a quantum analogue of classical phase-space partitioning~\cite{VonNeumann:10,Busch:93}. Similar techniques have been recently used to define coarse-grained or ``observational" entropies~\cite{Safranek:19,Safranek2:19,Buscemi23,Strasberg24}, with applications in the statistical mechanics description of equilibration in isolated quantum systems~\cite{Nagasawa:24,Schindler25,Meier25}. 

In this work, we focus on the role of coarse-graining in the thermodynamics of driven dissipative processes as seen from the point of view of modern information thermodynamics~\cite{Parrondo2015,Goold_2016}. Crucially, we show that, while finite measurement resolution can significantly alter the apparent properties of a process (such as work distributions, entropy production, and effective dynamics), the fundamental thermodynamic laws remain invariant under changes in this resolution. As an example, coarse-graining may affect the estimation of a process's degree of reversibility, making an irreversible process appear reversible (or vice versa), yet the average coarse-grained dissipative work required to transfer the system between coarse-grained thermal states always exceeds or equals their free energy difference. This leads to a family of second-law-like relations valid for any level of coarse-graining. In what follows, we first establish quantitative relations between work costs, coarse-graining, and entropy production in operational terms, linking coarse-graining to information thermodynamics. We then introduce a two-point measurement scheme scenario to derive coarse-grained counterparts of non-equilibrium relations, including the connection between dissipation and distinguishability from time-reversed processes~\cite{Kawai2007,Parrondo2009}, as well as quantum work fluctuation theorems~\cite{CampisiREV,EspositoREV}. From this, we conclude that every agent arrives at the same form of the fundamental thermodynamic laws, regardless of observational precision, even though the thermodynamic quantities involved depend on that precision.

The rest of this paper is organized as follows. Sec.~\ref{sec:CGThermalStates} introduces the coarse-graining procedure employed in this paper and the corresponding thermal states by applying the principle of maximum entropy to the case of a state estimated via limited precision measurements. 
Secs.~\ref{sec:Dissipative}-\ref{sec:FTs} focus on dissipative work due to driving protocols. In particular, we introduce a two-point-measurement scheme based on coarse-grained measurements  (Sec.~\ref{sec:Dissipative}), and derive second-law-like inequalities (Sec.~\ref{sec:CGSecondLaw}) and work fluctuation theorems (Sec.~\ref{sec:FTs}) for any resolution of the measurement apparatus. Sec.~\ref{sec:Relation_to_Other_CG} discusses the relation between the type of coarse-graining here considered and that in other previous works (\textit{e.g.}, Refs. \cite{Kawai2007,Parrondo2009}). Finally, Sec.~\ref{sec:GeneralizationToPOVM} extends the framework to non-orthogonal coarse graining, modelling readout noise and energy values assigned to different measurement outcomes, and discusses the implications for thermodynamic relations.

\section{Coarse-Grained Thermal States and Coarse-Graining Operation} 
\label{sec:CGThermalStates}

The tight link binding statistical mechanics and information theory has been highlighted in a number of seminal works \cite{Szilard1929, Landauer1961,Bennett1982}. One of the most relevant evidences of this interconnection is provided by the `principle of maximum entropy' \cite{Jaynes1957a,Jaynes1957b}. This principle states that the probability distribution which best represents the state of one's own knowledge is the one which leaves the greatest residual uncertainty (\textit{i.e.}, maximum entropy) consistent with given constraints (usually imposed by general conservation laws of macroscopic observables like energy, momentum or number of particles). However, in any practical investigation, the constraints entering the description of equilibrium states are themselves available only with finite precision, reflecting the limited resolution of the instruments or preparation procedures used to characterize the system.

Consider a physical system whose mean values of a number of physical observables are assigned based on experimentally accessible information with finite resolution. These observables could be, for instance, its energy, the number of particles composing the system, \textit{etc.}. Let us call $\langle O^{(k)} \rangle$ the mean values of $M$ physical observables of the system, with $k=1,\ldots,M$. The constraints on the mean values of such observables can be written as 
\begin{equation}\label{eq:observables}
\langle O^{(k)} \rangle = \sum_{i=1}^{N} p_i O^{(k)}_i,
\end{equation}
where $O^{(k)}_i$ are the possible values which each of the set of mutually commuting physical observable $\hat{O}^{(k)}$ can take in a measurement process, with probability $p_i$, that is, $\sum_{i=1}^N p_i =1$. Under these constraints, the probability distribution which exhibits maximum entropy (hence, the most likely according to the principle of maximum entropy) has the form:
\begin{equation}
\label{eqn:p_i}
 p_i  = \frac{1}{Z ( \beta_1 , \ldots , \beta_M) }\, \, \mathrm{exp}\Biggl(-{\sum_{k=1}^M \beta_k O^{(k)}_i}\Biggr),
\end{equation}
where $\beta_1, \ldots ,\beta_m$ are so-called `Lagrange multipliers', whose values are determined from the average values $\langle O^{(k)}\rangle= -\partial \log Z / \partial \beta_k$, and where the normalization constant $Z( \beta_1 , \ldots , \beta_M ) =\sum_{i=1}^{N} \mathrm{exp}\bigl(-{\sum_{k=1}^M \beta_k O^{(k)}_i}\bigr)$ is the partition function. (We note that the derivation can be extended to quantum statistical mechanics and to non-commuting observables \cite{Guryanova2016,YungerHalpern2016a}.)

One of the prime examples of the application of this principle is the derivation of the `thermal (Gibbs) state': 
\begin{equation}
\label{eqn:thermal}
{\tau}_{\beta} \equiv \frac{1}{Z} \sum_j e^{- \beta \epsilon_j} \vert \epsilon_j \rangle \langle \epsilon_j \vert = e^{-\beta (H-F)},
\end{equation}
with $Z = \sum_j e^{- \beta \epsilon_j}= e^{-\beta F}$ being the partition function of the system (where $F$ is the equilibrium free energy), $\beta = (k_B T)^{-1}$ being the inverse temperature, and $\lbrace \vert \epsilon_j \rangle \rbrace$ the energy eigenstates, with corresponding eigenvalues $\epsilon_j$ of the Hamiltonian $H$. This state maximizes the von Neumann entropy $S(\tau_\beta) = - \mathrm{Tr}\bigl[\tau_\beta \, \mathrm{ln}(\tau_\beta)\bigr]$ under the average energy constraint, or, equivalently, minimizes the average energy $\langle H \rangle = \tr[\tau_\beta H]$ for a fixed von Neumann entropy.

As a consequence, whenever one does not have access to the complete microscopic state of a system, the optimal state they shall attribute to it is the generalized Gibbs state in Eq.~\eqref{eqn:p_i} derived from the known mean values of the system observables~\cite{Jaynes1957a}. The link with thermodynamics is established through the notion of equilibrium~\cite{Callen_book}.
In other words, under interactions which preserve the relevant set of observables appearing in Eq.~\eqref{eqn:p_i}, the interaction between systems prepared in generalized Gibbs states does not produce any observable change in their states, nor any currents in the average values [Eq.~\eqref{eq:observables}]. This is particularly clear in the case of energy: systems in thermal contact may exchange heat with each other until they reach equilibrium states at the same temperature ${\tau}_{\beta}$.

\begin{figure}[tb]
\centering
\includegraphics[width=.65\columnwidth]{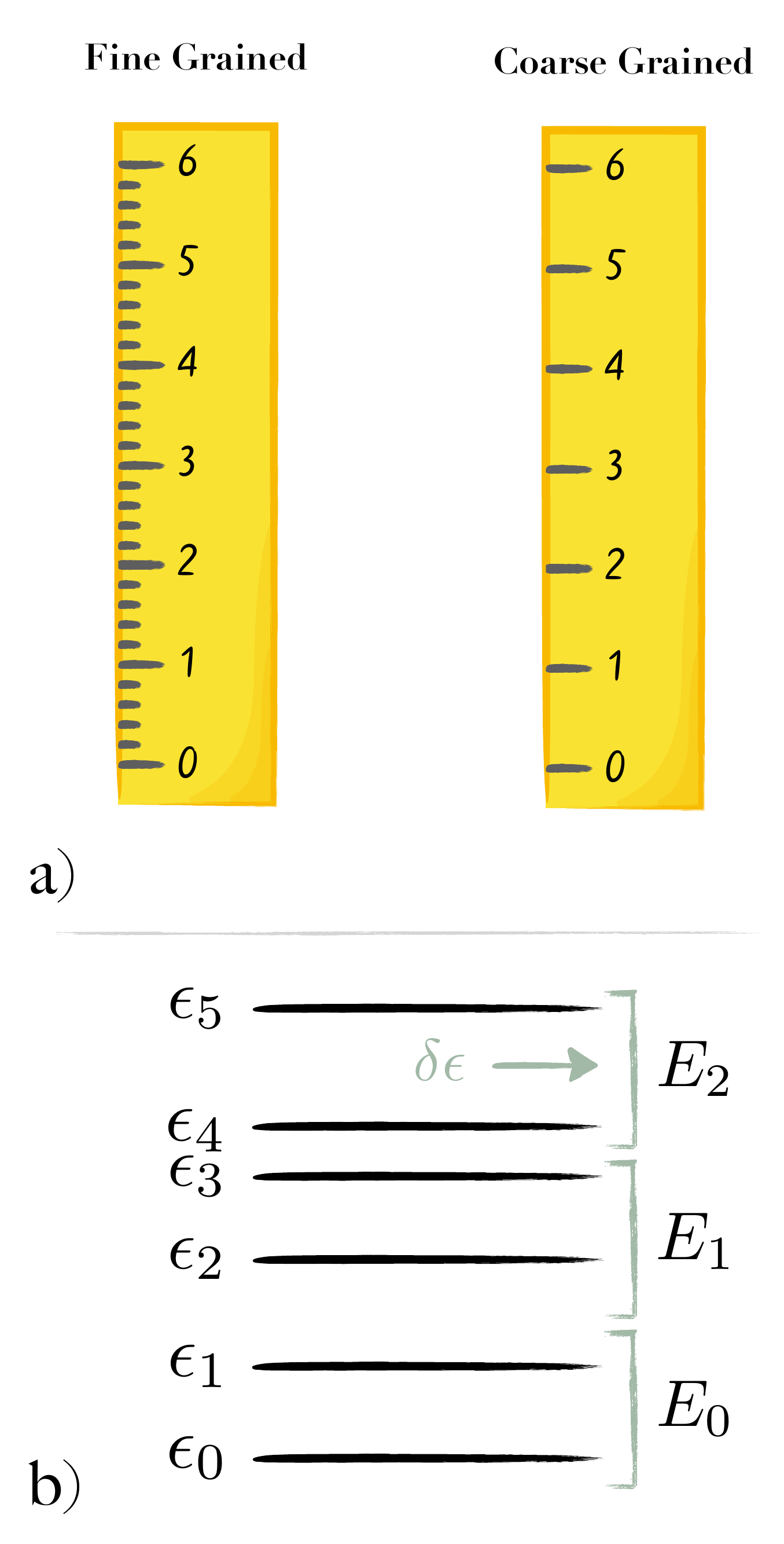}
\captionof{figure}{\footnotesize \textbf{Coarse-graining of a ruler and of the energy levels of a thermodynamic system.} \textbf{a)} A ruler shown without (left) and with (right) coarse-graining. Coarse-graining reduces the resolution of the instrument. Note: This is a schematic illustration (any instrument appears coarse-grained relative to one with higher precision, and only an idealised, infinitely precise instrument would be truly non-coarse-grained). \textbf{b)} Diagrammatic example illustrating how the levels $E_J$ in Eq.~\eqref{eq:E} are defined from the fine-grained energy levels $\epsilon_j$, using intervals of width $\delta \epsilon$.}
\label{img:CoarseGrainedRuler}
\end{figure}

In practice, however, not only does one not have direct access to the microscopic states of the system, but also the average values which one can evaluate are affected by the accuracy of the measuring instruments.
Performing a measurement under coarse-graining limits the ability to distinguish microscopic states of the system due to the limited resolution of the measurement device (see pictorial representation in Fig. \ref{img:CoarseGrainedRuler}a). By referring to the example in Fig. \ref{img:CoarseGrainedRuler}, let us label as $j$ the analogue of the `fine-grained' indentations of the ruler in the subfigure on the left, and as $J$ that of the `coarse-grained' indentations in the subfigure on the right.

Let $\{ {\Pi}_J \}$, with $\sum_J {\Pi}_J = {\mathbb{1}}$, be a set of non-zero projectors, which we call the coarse-graining projectors, and let us consider such a coarse-graining to be performed in the energy basis: 
\begin{equation}
{\Pi}_J = \sum_{j \in \mathcal{J}_J} \vert \epsilon_j \rangle \langle \epsilon_j \vert,
\end{equation}
with intervals $\mathcal{J}_J = \Bigl[\sum_{K = 1}^{J-1} \mathrm{Tr}[{\Pi}_K]+1, \sum_{K = 1}^{J} \mathrm{Tr}[{\Pi}_K] \Bigl]$ that we call `slots', each of them containing a number of microstates $N_J \equiv \tr[\Pi_J]$. Under such a coarse-graining, the original `fine-grained' Hamiltonian $H = \sum_j \epsilon_j \ket{\epsilon_j}\bra{\epsilon_j}$, becomes
\begin{equation} \label{eq:coarseH}
\breve{H} = \sum_{J=1}^{\breve{N}} E_J \Pi_J,
\end{equation}
with degenerate eigenvalues $E_J$, and where we labeled $\breve{N}$ the number of slots. We will refer to this observable as the `coarse-grained energy' (we consider $\breve{N} > 1$). 
Here, the intervals $\mathcal{J}_J$ define the coarse-grained slots. These, in principle, can be chosen arbitrarily. However, in this work we are particularly interested in a physically motivated prescription based on the energy resolution of the measuring instruments. Specifically, we choose the (fine-grained) ground state as a natural reference point to define the intervals\footnote{The ground state is chosen for convenience; any consistent reference point would work to prevent overlaps and ensure a valid projector decomposition.}, and define each slot $J$ to contain all eigenstates $\{ \ket{\epsilon_j} \}$ such that $ (J-1) \delta \epsilon \leq \epsilon_j < J \delta \epsilon$, where $\delta \epsilon$ denotes the smallest energy difference resolvable by the instruments, and $J = 1, 2, \ldots, \breve{N}$ (see Fig.~\ref{img:CoarseGrainedRuler}\textbf{b)} for an illustrative example)\footnote{This prescription is more stringent than what one might expect from a plain coarse-graining procedure, namely, grouping together any pair of eigenstates $\ket{\epsilon_j}$ and $\ket{\epsilon_l}$ whenever their energy difference satisfies $|\epsilon_j - \epsilon_l| < \delta \epsilon$. Such an approach may seem natural, but it fails to guarantee a valid projector decomposition. To see this, consider, e.g., the case of equally spaced energy levels with spacing $0.1\varepsilon$ and an energy resolution $\delta \epsilon = 0.15\varepsilon$, where $\varepsilon$ is an energy-scale parameter with dimensions of energy. In this scenario, levels 1 and 2 would be indistinguishable, as would levels 2 and 3, and so on, leading to transitive overlaps across many levels. As a result, the same fine-grained state could end up being included in more than one coarse-grained slot. This overlap results in the projectors $\Pi_J$ failing to add up to the identity, as $\sum_J \Pi_J > \mathbb{1}$, and thus they cannot define a consistent coarse-graining. Our interval-based construction avoids this inconsistency by assigning each energy eigenstate to exactly one slot.}.

Given the above constraint, and assuming energy is the only relevant conserved quantity, the maximum entropy principle yields what we will call the `coarse-grained thermal state':
\begin{equation}
\label{eqn:expRho}
\breve{\tau}_\beta \equiv e^{-\beta(\breve{H} - \breve{F})} = \sum_{J} \frac{e^{- \beta E_J}}{\breve{Z}} \Pi_J,
\end{equation}
where $\breve{Z} \equiv e^{-\beta \breve{F}} = \sum_{J} e^{- \beta E_J} N_J$ is the coarse-grained partition function. This form follows from applying the maximum entropy principle under the constraint that only coarse-grained energy measurements are accessible: since measurement outcomes correspond to projectors ${\Pi_J}$, the optimal state must be block-diagonal in this basis, with weights proportional to $e^{-\beta E_J}$ within each slot. One should note, however, that the specific values $E_J$ in Eq.~\eqref{eqn:expRho} are not determined by the maximum entropy principle alone, but need to be derived from further assumptions ensuring thermodynamic consistency, as we will shortly see. Coarse-grained thermal states in this form provide the central mathematical tool for the analysis developed in this work.

In order to introduce a link between coarse-grained and fine-grained pictures, we use the coarse-graining mapping $\mathcal{C}_{\Pi_J}(\rho) \rightarrow \breve{\rho}$, associated to a set of coarse-graining projectors $\{\Pi_J\}$, which transforms any state in the fine-grained description to the coarse-grained one:
\begin{equation} \label{eq:cgmap}
\mathcal{C}_{\Pi_J}(\rho) \equiv \sum_J \frac{\tr[\rho~ \Pi_J]}{N_J} \Pi_J = \sum_J \breve{p}_J \frac{\Pi_J}{N_J},   
\end{equation}
where $\breve{p}_J$ denotes the coarse-grained eigenvalues of $\breve{\rho} \equiv \mathcal{C}_{\Pi_J}(\rho)$, with $\sum_J \breve{p}_J = 1$ (see also Ref.~\cite{Buscemi23}). Notice also the need to keep the factor $N_J$ in the denominator since $\tr{\Pi_J}=N_J$. Loosely speaking, the role of the map $\mathcal{C}_{\Pi_J}$ is essentially to flatten the density operator $\rho$ within the slots defined by the projectors $\Pi_J$, that is, it replaces each block $J$ of the fine-grained state
by a flat diagonal block proportional to the maximally mixed state $\mathbb{1}_J$.

The map $\mathcal{C}_{\Pi_J}$ captures the fact that coarse-graining leads to a loss of the microscopic information about the state $\rho$, thereby maximizing entropy within the slots inaccessible to the coarse-grained instruments. In particular, $\mathcal{C}_{\Pi_J}(\rho + \delta \rho) = \mathcal{C}_{\Pi_J}(\rho)$ for any perturbation  $\delta \rho$ such that $\tr[\delta \rho \; \Pi_J] = 0$ for all $J$, i.e., for any $\delta \rho$ that does not modify the overall population of the slots $J$. Consequently, the map $\mathcal{C}_{\Pi_J}$ corresponds to a logically irreversible operation, which transforms fine-grained states with potentially different microscopic structures to the same (block-flattened) density $\breve{\rho}$. Importantly, this description does not depend on how the slots are chosen. In particular, we note that $\{\Pi_J\}$ may in principle represent any projective partition of the energy eigenbasis; the derivations in the orthogonal coarse-graining setting rely only on the fact that the associated map $\mathcal{C}_{\Pi_J}$ is completely positive, trace preserving, and unital, and therefore do not depend on whether the sets $\mathcal{J}_J$ correspond to contiguous energy intervals.

From a different but related viewpoint, one may also ask whether the coarse-grained description should be formulated on a reduced Hilbert space. Although the coarse-grained state $\breve{\rho}$ can be equivalently represented on a reduced Hilbert space of dimension $\breve{N}$, since each block $\Pi_J \breve{\rho} \Pi_J$ is proportional to the identity within the corresponding slot, we keep the full Hilbert space throughout. This is convenient because the unitary dynamics implementing the driving protocol is naturally defined on the fine-grained space and may, in general, act nontrivially within each slot. Retaining the full Hilbert space therefore allows us to describe consistently both the coarse-grained state assignment and the underlying physical dynamics within a single framework.

On the more technical side, $\mathcal{C}_{\Pi_J}$ is a completely-positive and trace-preserving (CPTP) map, which can be written in terms of Kraus operators as $\mathcal{C}(\rho) = \sum_k M_k \rho M_k^\dagger$ fulfilling $\sum_k M_k^\dagger M_k = \mathbb{1}$. This map preserves any state already flattened within the slots, which are all invariant states of the map. It is also an idempotent operation, \textit{i.e.} $\mathcal{C}_{\Pi_J}^2(\rho) = \mathcal{C}_{\Pi_J}(\rho)$, and it preserves the identity operator, $\mathcal{C}_{\Pi_J} (\mathbb{1}) = \mathbb{1}$, that is, it is a \emph{unital} map. It hence verifies $S(\mathcal{C}_{\Pi_J}(\rho)) \geq S(\rho)$ for any $\rho$. Another important property of the map $\mathcal{C}_{\Pi_J}$ is that it preserves the coarse-graining energy $\tr[\breve{H} \rho] = \tr[\breve{H} \mathcal{C}_{\Pi_J}(\rho)]$ for any state $\rho$. In this sense, the coarse-grained state is an effective description of the system. We also notice that the convolution of the map $\mathcal{C}_{\Pi_J}$ with von Neumann entropy: 
\begin{equation}
\begin{split}
S\bigl(\mathcal{C}_{\Pi_J}(\rho)\bigr) =& \sum_J \sum_{j \in \mathcal{J}_J} \left(\frac{\breve{p}_J}{N_J}\right)\log\left(\frac{\breve{p}_J}{N_J}\right) \\ =& \sum_J \breve{p}_J \log\left(\frac{\breve{p}_J}{N_J}\right),        
\end{split}
\end{equation}
naturally leads to the definition of coarse-grained or \emph{observational} entropies~\cite{Safranek:19,Safranek2:19,Buscemi23,Nagasawa:24,Schindler25}.

We now apply the coarse-graining map $\mathcal{C}_{\Pi_J}$ in Eq.~\eqref{eq:cgmap} to the thermal state $\tau_\beta$. The result of this operation must correspond to the coarse-grained thermal state $\breve{\tau}_\beta$ in Eq.~\eqref{eqn:expRho} within the precision of the instruments, that is:
\begin{equation}\label{eq:condition}
\mathcal{C}(\tau_\beta + \delta \tau) = \breve{\tau}_\beta,    
\end{equation}
where $\delta \tau$ is a (fine-grained) perturbation such that $\tr[\delta \tau \Pi_J] = 0$ for all $J$. This consistency condition is key to determining the coarse-grained energy values $E_J$ in Eq.~\eqref{eq:coarseH} and, as we see in Appendix~\ref{sec:thermocost}, it is in fact implied by the second law of thermodynamics. 
Applying the preservation of the coarse-graining Hamiltonian to the thermal state $\tau_\beta$, and using the condition in Eq.~\eqref{eq:condition}, we obtain\footnote{Note that this relation can equivalently be derived directly from $\mathcal{C}_{\Pi_j}(\tau_\beta) = \breve{\tau}_\beta$, without invoking Eq.~\eqref{eq:condition}, which is more general and accounts for a generic variation $\delta \tau$, explicitly showcasing the logically irreversible character of coarse-graining.}:
\begin{equation}\label{eq:condition2}
e^{- \beta E_J} ~N_J = \sum_{j \in \mathcal{J}_J} e^{- \beta \epsilon_j},
\end{equation}
which also implies $e^{-\beta \breve{F}} = \sum_{J} e^{- \beta E_{J}} N_J = \sum_{j} e^{- \beta \epsilon_{j}} = e^{-\beta F}$, or, equivalently, $\breve{F} = F$.
Based on Eq.~\eqref{eq:condition}, each coarse-grained energy level takes on a value
\begin{equation} \label{eq:E}
E_J = -k_B T \, \log \left(\frac{Z_J}{N_J}\right),
\end{equation}
where by $Z_J \equiv \mathrm{Tr} [e^{- \beta H} \Pi_J]$ we denoted the partition function within the slot $J$. Notably, the coarse-grained eigenvalues $E_J$ depend not only on the energy eigenvalues of the original Hamiltonian $H$, but also on the temperature of the reservoir $T$. 

In Appendix~\ref{app:B}, we show that coarse-grained eigenvalues $E_J$ in Eq.~\eqref{eq:E} lie in the interval between the minimum eigenvalue in the slot, $E_J^{\min} \equiv \min_{j \in \mathcal{J}_J} \epsilon_j$, and their average:
\begin{equation}
\label{eq:boundsE}
E_J^{\min} \leq E_J \leq \bar{E}_J,
\end{equation}
where $\bar{E}_J = \sum_{j \in \mathcal{J}_J} \epsilon_j/N_J$ is the slot mean. We notice that these two limiting cases correspond to the low- and high-temperature limits, respectively: For low temperatures $Z_J/N_J \sim e^{-\beta E_J^{\min}}$, and hence $E_J \sim E_J^{\min}$. On the other hand, for high temperatures ($\beta \epsilon_i \ll 1$), we can series-expand the exponential term, thereby obtaining $Z_J \simeq N_J(1 -\beta \bar{E}_J)$, which implies $E_J \simeq \bar{E}_J$ from Eq.~\eqref{eq:E}.  
The bounds in Eq.~\eqref{eq:boundsE} guarantee that the energy eigenvalues appear distinguishable to coarse-grained instruments, i.e., $|E_J - E_K| > \delta \epsilon$ for any pair of slots $K,J$, and that they do not intersect (analogously to the case of the reduction in resolution of a ruler in Fig.~\ref{img:CoarseGrainedRuler}).

Finally, the prescription introduced above specifies a thermal state for any size of coarse-grained slots, which are determined by the accuracy of the agent's instruments to measure energies. Each such range defines a different coarse-grained energy for a given temperature. In the upcoming sections, the state in Eq.~\eqref{eqn:expRho} will be adopted as the basis for a formulation of the thermodynamic laws for coarse-grained thermal states. 

\section{Dissipative work and fluctuations}
\label{sec:Dissipative}

The information-theoretic measure of the quantum relative entropy between the states subject to a (generally irreversible) thermodynamic process is directly related to the average dissipated work during the process~\cite{Maes:2003,Kawai2007,Parrondo2009,Horowitz:2009,Sagawa,Gaveau:2014}. Moreover, fluctuations in dissipative work feature universal properties characterised by a set of non-equilibrium equalities conventionally referred to as fluctuation theorems~\cite{JarzynskiREV,CampisiREV,EspositoREV}. Here, we are interested in the dissipative work in processes described \emph{and performed} by both fine-grained and coarse-grained agents. In the following, we first review the prototypical scenario corresponding to a fine-grained agent $F$, and then introduce the key modifications in the case of the coarse-grained agent $C$. Importantly, in our setup, the different level of resolution has implications in the way an experiment is prepared and not only observed.

\subsection{Two-point measurement scheme and thermodynamic relations for a fine-grained agent}
 
 Consider a quantum system with time-dependent Hamiltonian $H(t)$ which is initially prepared in a thermal state $\tau_\beta^0$, to undergo unitary evolution within which the Hamiltonian of the system is changed from $H_0 \equiv H(0)$ to $H_\tau \equiv H(\tau)$ in a time window $t \in (0, \tau]$ by changing some control parameter of the Hamiltonian, $\lambda(t)$, according to a prescribed protocol $\Lambda = \{\lambda(t); 0\leq t \leq \tau \}$. The dynamical evolution occurs in isolation from the thermal environment~\footnote{This typically happens when the dynamics are much faster than the relaxation timescale of the system in contact with the environment.}, so that it can be described by a unitary operator $U(t,0) \equiv \mathcal{T} e^{-i \int_0^t H[\lambda(s)] ds}$, where $\mathcal{T}$ denotes the time-ordering operator. At the end of the process, the system is generally in a non-equilibrium state, $\rho(\tau) = U(\tau,0) \tau_\beta^0 U(\tau,0)^\dagger$, which then interacts with the thermal environment and relaxes to the thermal state corresponding to the final Hamiltonian, $\tau_\beta^\tau$. The dissipated work in the process, $W_{\text{diss}} = W - \Delta F$, is defined as the extra amount of work which has been invested in the transformation, $W$, on the top of the free-energy difference, $\Delta F$, between the initial and final equilibrium states.
 
 The average dissipated work in the process is proportional to the quantum relative entropy between the states of the system in forward and time-reversal processes at any intermediate instant of time~\cite{Kawai2007, Parrondo2009}:
\begin{equation}
\label{eqn:W_diss}
\langle W\rangle - \Delta F = k_B T~ S(\rho(t) \vert \vert \Theta^\dagger \tilde{\rho}(\tau - t) \Theta),
\end{equation}
where $\langle W \rangle = \tr[H_\tau \rho(\tau)] - \tr[H_0 \tau_\beta^0]$ is the average work performed during the implementation of protocol $\Lambda$, and we introduced the system states $\rho(t) = U(t,0) \tau_\beta^0 U(t,0)^\dagger$ and $\tilde{\rho}(\tau - t) = \tilde{U}(\tau -t,0) \, \Theta\tau_\beta^\tau \Theta^\dagger \, \tilde{U}(\tau -t,0)^\dagger$ at intermediate times $t \in [0, \tau]$ in forward and time-reversal process, respectively. The time-reverse process is defined by the application of the (reversed) driving protocol $\tilde{\Lambda} = \{\lambda(\tau - t); 0\leq t \leq \tau \}$ over the final thermal state $\tau_\beta^\tau$, as described by the unitary $\tilde{U}(t,0)=\mathcal{T} e^{-i \int_0^t \Theta H[\lambda(\tau - s)] \Theta^\dagger ds}$. In the above equations, $\Theta$ denotes the time-reversal operator in quantum mechanics, which reverses the sign of observables with odd parity under time reversal. We also emphasise that the two density operators appearing in Eq.~\eqref{eqn:W_diss} are evaluated at the same point in time of their respective evolution\footnote{Here we assumed for simplicity that the Hamiltonian of the system is invariant under time-reversal at initial and final times, that is, $[H_0, \Theta]= [H_\tau, \Theta] = 0$. However, this assumption could be avoided by considering as the initial condition for the time-reversed process, the inverted thermal state $\Theta \rho_\beta^\tau \Theta^\dagger$.}. As a result, the dissipative work can be interpreted as quantifying the distinguishability between the forward and time-reversed processes. In this sense, it serves as a measure of how effectively one can infer the arrow of time from the observed physical process.

The positivity of the quantum relative entropy in Eq.~\eqref{eqn:W_diss} is a manifestation of the second law of thermodynamics, that is, the average external work required to perform protocol $\Lambda$ cannot be smaller than the system's free-energy change, $\langle W\rangle \geq \Delta F$. Stronger thermodynamic constraints also appear in the setup by taking into account work fluctuations. These fluctuations can be accessed via projective measurements of the system Hamiltonian at the beginning and at the end of the unitary evolution, a framework which has been dubbed two-point measurement (TPM) scheme~\cite{CampisiREV,EspositoREV}. 

In the TPM scheme, a first projective measurement of $H_0$ is performed on the initial state $\tau_\beta^0$, obtaining an outcome $n$ with probability $p_n^{(0)}= e^{-\beta \epsilon_n^{(0)}}/Z_0$ and corresponding eigenvalue $\epsilon_n^{(0)}$. Then, after the implementation of the protocol $\Lambda$ through the unitary $U(\tau,0)$, a second projective measurement of $H_\tau$ is performed, giving outcome $m$ with eigenvalue $\epsilon_m^{(\tau)}$. The very same procedure can also be applied to the time-reversed process by exchanging the order of the measurements and replacing $U(\tau,0)$ by $\tilde{U}(\tau, 0)$.

Combining the statistics of measurements outcomes in many realizations of the TPM scheme, one can construct the work probability distribution associated to the forward protocol $\Lambda$ as $P(W) = \sum_{n m} P_{n,m} \delta(W - W_{n,m})$, where $P_{n,m} = p_n^{(0)} |\bra{\epsilon_m^{(\tau)}} U(\tau,0) \ket{\epsilon_n^{(0)}}|^2$ is the joint probability to obtain outcomes $n$ in the first measurement and $m$ in the second one, and $\delta(x)$ denotes the Dirac delta function with $W_{n,m} \equiv \epsilon_m^{(\tau)} - \epsilon_n^{(0)}$. Analogously, the work probability distribution in the time-reverse process reads $\tilde{P}(W)=\sum_{n, m} \tilde{P}_{n m} \delta(W + W_{n,m})$, with $\tilde{P}_{n m} = p_m^{(\tau)} |\bra{\epsilon_n^{(0)}} \Theta^\dagger \tilde{U}(\tau,0) \Theta  \ket{\epsilon_m^{(\tau)}}|^2$ being the joint probability to obtain, correspondingly, outcomes $m$ and $n$ before and after the application of protocol $\tilde{\Lambda}$. These two probability distributions verify the detailed fluctuation theorem~\cite{CrooksTheorem,CampisiREV}
\begin{equation} \label{eq:crooks}
\frac{P(W)}{\tilde{P}(-W)} = e^{\beta(W - \Delta F)},
\end{equation}
which provides a link between irreversibility (through the probabilities of work in forward and time-reverse processes), and the dissipative work, in every realization of the TPM. The fluctuation theorem in Eq.~\eqref{eq:crooks}, together with the integral version $\langle e^{-\beta (W - \Delta F)} \rangle = 1$, can be regarded as deeper manifestations of the second law, putting constraints over the structure of  fluctuations of thermodynamic quantities~\cite{JarzynskiREV}.

\subsection{Coarse-grained TPM scheme}

\begin{figure}[tb]
\centering
\includegraphics[width=\columnwidth]{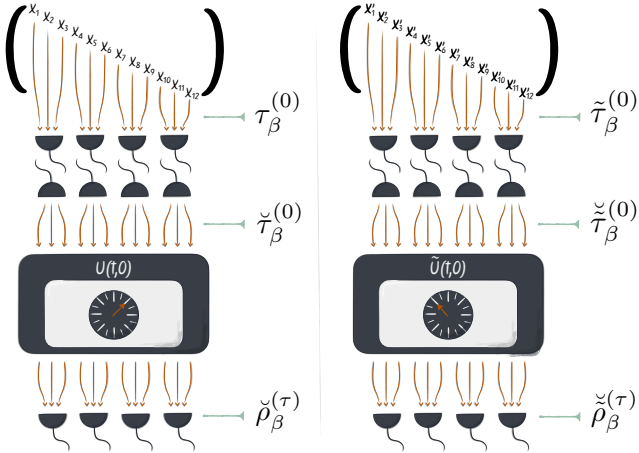}
\captionof{figure}{\footnotesize
\textbf{Schematic of the operations performed on the system in the forward (left) and time-reversed (right) processes.} A thermodynamic system is initially prepared in the (fine-grained) state ${\tau}^{(0)}_{\beta}$ in the forward process, or in $\tilde{\tau}^{(0)}_{\beta} = \Theta \tau^{(\tau)}_\beta \Theta^\dagger$ for the time-reversed process. It is then measured and re-prepared using coarse-grained instruments, yielding the state $\breve{\tau}^{(0)}_{\beta}$ (or $\breve{\tilde{\tau}}^{(0)}_{\beta} = \Theta \breve{\tau}^{(\tau)}_\beta \Theta^\dagger$ for the time reversal) at time $t=0$. Subsequently, the system evolves under a unitary evolution $U(t,0)$ (or $\tilde{U}(t,0)$ in the time-reversed case) during the interval $t \in (0, \tau]$, reaching the final state $\breve{\rho}_{\beta}^{(\tau)} = U(\tau,0) \, \breve{\tau}^{(0)}_{\beta} \, U^\dagger(\tau,0)$ at $t=\tau$ ($\breve{\tilde{\rho}}_{\beta}^{(\tau)} = \tilde{U}(\tau,0) \, \breve{\tilde{\tau}}_{\beta}^{{(0)}} \, \tilde{U}^\dagger(\tau,0)$ at $t=\tau$ for the time-reversed process).}
\label{img:drawing}
\end{figure}

We now extend the validity of Eqs. \eqref{eqn:W_diss} and \eqref{eq:crooks} to coarse-grained quantities. 
Let us first specify how exactly the lack of resolution of the instruments of the coarse-graining  agent $C$ affects the standard TPM scheme (see Fig.~\ref{img:drawing}). Our starting point is again the (fine-grained) thermal state $\tau_\beta^{(0)}$, over which a first projective energy measurement of $\breve{H}_0$ is performed. In this case, agent $C$ obtains outcome $I$, corresponding to eigenvalue $E_I^{(0)}$, with probability
\begin{equation}
    \breve{p}_I^{(0)} = \tr[\tau_\beta^{(0)} \Pi_I^{(0)}] = \dfrac{e^{-\beta E_I^{(0)}}}{Z_0} N_I^{(0)},
\end{equation}
where we used Eq.~\eqref{eq:condition2}. After the measurement, agent $C$ \emph{reprepares} the state $\Pi_I^{(0)}/N_I^{(0)}$, and applies the unitary evolution $U(\tau,0)$ associated with protocol $\Lambda$, with $U(\tau,0)$ potentially acting on fine-grained degrees of freedom that are not accessible to the coarse-graining agent $C$\footnote{This modelling choice is intentional and reflects the fact that the physical evolution of a system need not be constrained by the measurement resolution of a given agent. Even when an agent can only resolve a restricted set of observables, the underlying dynamics may still involve fine-grained degrees of freedom, for instance due to natural evolution, external driving, or operations implemented by another agent or automated mechanism. Adopting the same fine-grained unitary evolution in both the fine- and coarse-grained TPM schemes therefore allows for a consistent comparison that isolates the effects of limited measurement resolution and state re-preparation from those of the underlying dynamics.}. 
At the end of the protocol, a second measurement of $\breve{H}_\tau$ is performed, obtaining the outcome $J$ associated with the eigenvalue $E_J^{(\tau)}$. Finally, the system is put again in contact with the thermal reservoir until it equilibrates with the fine-grained state $\tau_\beta^{(\tau)}$ corresponding to the final Hamiltonian $H_\tau$.

The joint probability of obtaining outcomes $I$ and $J$ in the coarse-graining process then reads
\begin{equation}
\label{eq:P_IJ}
   \breve{P}_{I, J} = \breve{p}_{I}^{(0)}\tr\Biggl[\Pi_J^{(\tau)} U(\tau,0)~ \dfrac{\Pi_I^{(0)}}{N_I^{(0)}} U^\dagger(\tau,0)\Biggr]. 
\end{equation}
It is worth noting that the TPM procedure with coarse-grained instruments effectively involves a repreparation of the initial fine-grained thermal state into the corresponding coarse-grained thermal state. Specifically, the ensemble average over the post-measurement states yields
\begin{align}
\breve{\tau}_{\beta}^{(0)} &= \sum_I \mathrm{Tr}[\tau_\beta^{(0)} \Pi_I^{(0)}] \, \frac{\Pi_I^{(0)}}{N_I^{(0)}} = \mathcal{C}_{\Pi_I^{(0)}}(\tau_\beta^{(0)}),
\end{align}
which is precisely the result of applying the coarse-graining operation defined in Eq.~\eqref{eq:cgmap}. This stands in contrast to a procedure in which the initial state $\tau_\beta^{(0)}$ is merely \emph{projected} onto the $I$-th subspace, yielding $\Pi_I^{(0)} \tau_\beta^{(0)} \Pi_I^{(0)} / \breve{p}_I^{(0)}$, which retains the fine-grained structure of the probability distribution within the slot $I$. We observe that, while these two procedures appear to have the same effect from the perspective of a coarse-graining agent C, they affect the underlying fine-grained degrees of freedom and subsequent dynamics (i.e., unitary evolution), as observed by any agent.

The time-reversed protocol is defined as follows. It begins with the (fine-grained) thermal state $\tau_\beta^{(\tau)}$, upon which a measurement of the coarse-grained Hamiltonian $\breve{H}_\tau$ is performed, yielding outcome $J$. As in the forward process, this measurement is accompanied by the repreparation of the state $\Pi_J^{(\tau)}/N_J^{(\tau)}$. At the ensemble level, this is equivalent to applying the coarse-graining operation to the initial state, resulting in
\begin{align}
\breve{\tau}_{\beta}^{(\tau)} &= \sum_J \mathrm{Tr}[\tau_\beta^{(\tau)} \Pi_J^{(\tau)}] \, \frac{\Pi_J^{(\tau)}}{N_J^{(\tau)}} = \mathcal{C}_{\Pi_J^{(\tau)}}(\tau_\beta^{(\tau)}).
\end{align}
Subsequently, the unitary evolution $\tilde{U}$ associated to protocol $\tilde{\Lambda}$ is applied, after which the final measurement of the coarse-grained Hamiltonian $\breve{H}_0$ is performed, obtaining outcome $I$. Therefore, the joint probability of obtaining outcomes $J$ and $I$ in the time-reversed (coarse-grained) process reads
\begin{equation}
    \breve{\tilde{P}}_{I, J} = \breve{p}_{J}^{(\tau)}\tr\Biggl[\Pi_I^{(0)} \tilde{U}(\tau,0)~ \dfrac{\Pi_J^{(\tau)}}{N_J^{(\tau)}} \tilde{U}^\dagger(\tau,0)\Biggr],
\end{equation}
where $\breve{p}_{J}^{(\tau)} = \mathrm{Tr}[\tau_\beta^{(\tau)} \Pi_J^{(\tau)}]= \frac{e^{-\beta E_J^{(\tau)}}}{Z_\tau} N_J^{(\tau)}$.

The coarse-grained TPM scheme introduced here provides a systematic way to describe thermodynamic processes under limited measurement resolution, making explicit how finite precision affects both the statistics of observed work and the reconstruction of the system’s state. In the following, we illustrate these concepts through a concrete example involving a driven 12-level system.

\subsection{Case study: A driven 12-levels system}
\label{subsec:Case_Study}

\begin{figure}[tb]
\centering
\includegraphics[width=.85\columnwidth]{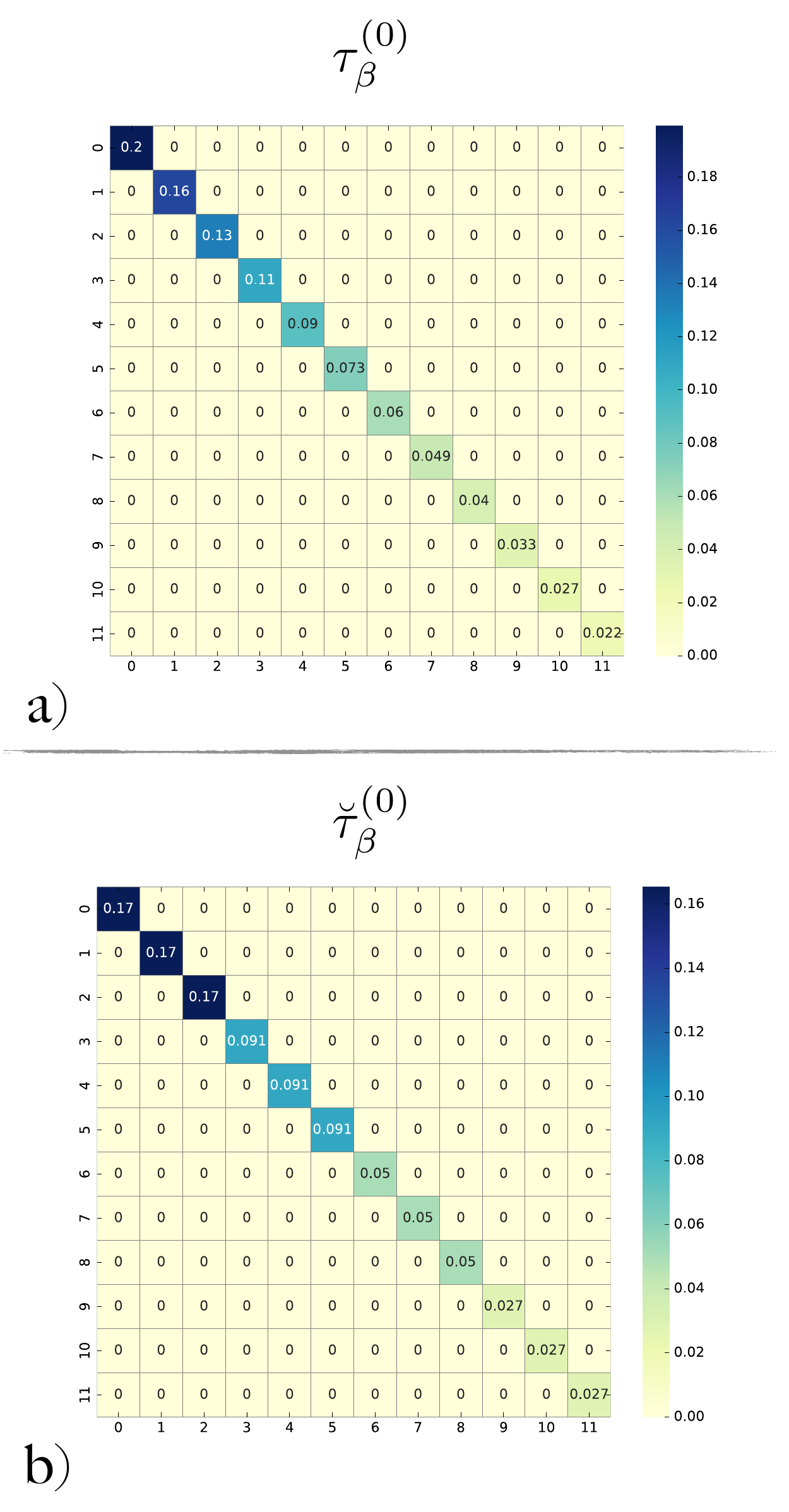}
\captionof{figure}{\footnotesize \textbf{Matrix elements of fine- and coarse-grained density matrices of a harmonic oscillator.} Fine-grained (Panel \textbf{a)}) and coarse-grained (Panel \textbf{b)}) density matrices of the thermal states $\tau_\beta^{(0)}$ and $\breve{\tau}_\beta^{(0)}$, respectively, corresponding to a 12-levels harmonic oscillator. The system's Hamiltonian is $H_0 = \hbar \omega \, (a^\dagger a +\mathbb{1}/2)$, having set $\omega = 2$ and $\hbar = 1$. The thermal state $\breve{\tau}_\beta^{(0)}$ is defined as in Eq. \eqref{eqn:thermal}, with $\epsilon_j^{(0)} = \hbar \omega (j + 1/2)$ being the energy levels of the harmonic oscillator, and $\beta = (k_B T)^{-1}$, with $k_B = 1$, $T=10$. Here, each coarse-grained interval is composed of a $3 \times 3$ fine-grained submatrices (i.e., $\mathrm{Tr}[P_J] = 3$).}
\label{img:rho0_FineAndCoarseGrained}
\end{figure}

As an illustrative example, let us consider a 12-level harmonic oscillator of Hamiltonian $H_0 = \hbar \omega \, (a^\dagger a +\mathbb{1}/2)$, with $a$ and $a^\dagger$ being the annihilation and creation operators such that $a \vert n\rangle = \sqrt{n} \vert n-1\rangle$, $a^\dagger \vert n\rangle = \sqrt{n+1} \vert n+1\rangle$, and $\omega=2$, $\hbar=1$. Imagine that a coarse-grained measurement is carried out on such a system, and that this measurement is only able to resolve a minimum of, say, three energy levels\footnote{Here, we describe the resolution in terms of energy levels rather than an absolute energy scale $\delta \epsilon$ because the harmonic oscillator has equally spaced levels. In this case, an inability to resolve fewer than three energy levels automatically corresponds to a well-defined $\delta \epsilon$.}, \textit{i.e.}, $\mathcal{C} = \{P_1, P_2, P_3, P_4\}$ with ${P}_1 = \sum_{j=1}^3 \vert \epsilon_j \rangle \langle \epsilon_j \vert$, ${P}_2 = \sum_{j=4}^6 \vert \epsilon_j \rangle \langle \epsilon_j \vert$, ${P}_3 = \sum_{j=7}^9 \vert \epsilon_j \rangle \langle \epsilon_j \vert$, and ${P}_4 = \sum_{j=10}^{12} \vert \epsilon_j \rangle \langle \epsilon_j \vert$. Fig.~\ref{img:rho0_FineAndCoarseGrained} shows the comparison between the fine-grained density matrix of the system $\tau^{(0)}_{\beta}$, and its coarse-grained counterpart $\breve{\tau}_\beta^{(0)}$.

Suppose that, immediately after having measured and reprepared the system at $t = 0$, an agent turns on a time-independent driving force $f$ acting on the system in the state $\breve{\tau}_\beta^{(0)}$. The Hamiltonian, for $t \in (0, \tau]$, takes the form $H_\tau = \hbar \omega \, (a^\dagger a +\mathbb{1}/2) + f (a^\dagger + a)/\sqrt{2}$, which does not commute with the initial Hamiltonian and therefore induces transitions between energy eigenstates\footnote{For comparison, in App.~\ref{app:C} we analyze an alternative driving protocol that commutes with $H_0$, corresponding to a purely classical (number-conserving) evolution.}.

\begin{figure}[bt]
\centering
\includegraphics[width=\columnwidth]{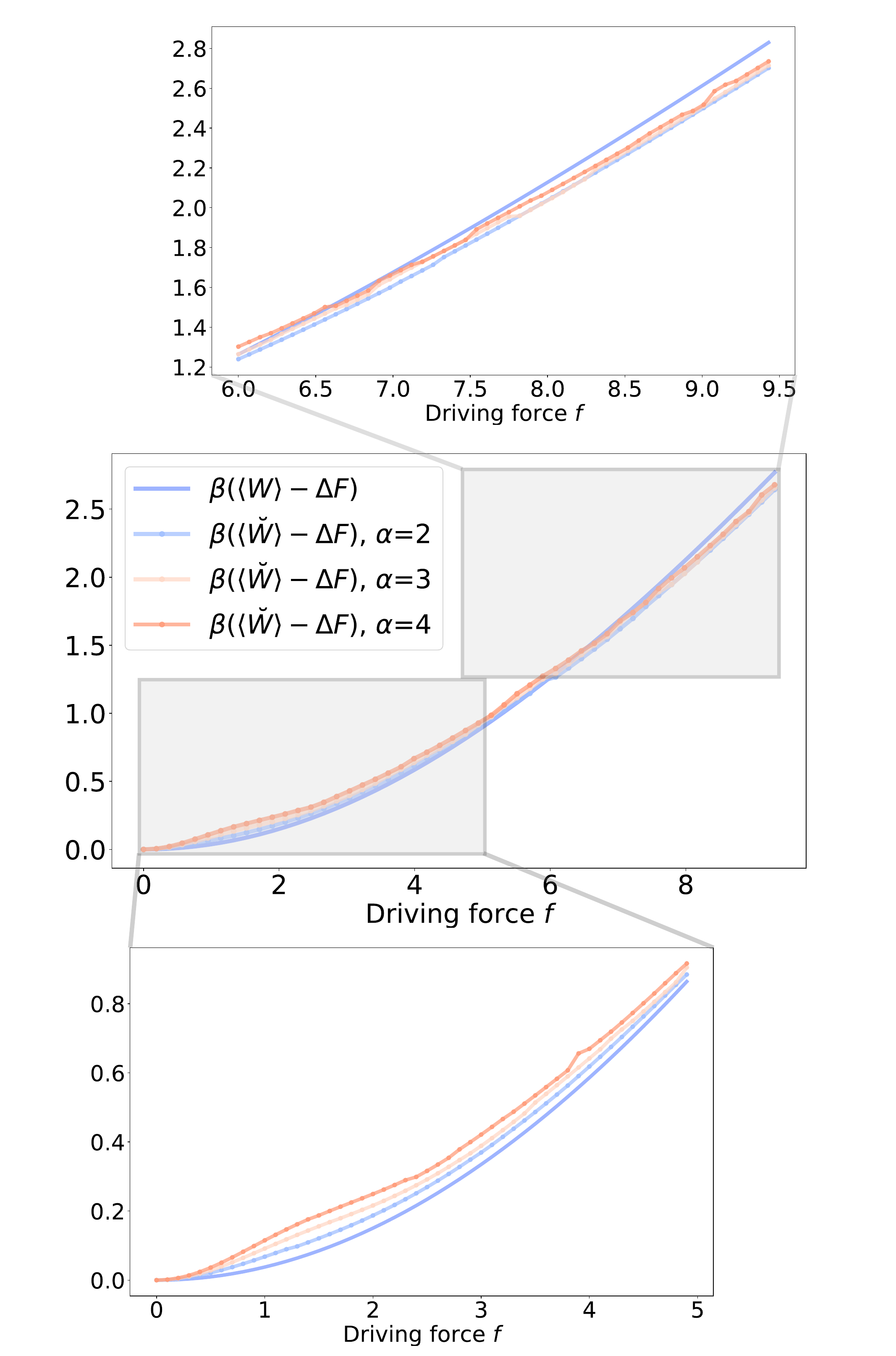}
\captionof{figure}{\footnotesize Trends of fine- and coarse-grained dissipative work for a harmonic oscillator under a time-independent driving force $f \in [0,9.5]$. The initial state is as in Fig.~\ref{img:rho0_FineAndCoarseGrained}, and the final state is prepared by applying a constant force $f$ within the interval $t \in (0,\tau]$, with Hamiltonian $H_\tau = \hbar \omega (a^\dagger a + 1/2) + f(a^\dagger + a)/\sqrt{2}$ (with $f$ expressed in energy units, and $\hbar = 1$). The plots show fine-grained and coarse-grained dissipated work (relative entropy) and average work as functions of $f$, for fixed $\tau = 1$, $\beta = 1$ and $\omega = 2$. (Full details on the case study considered are provided in Sec.~\ref{subsec:Case_Study}.) Coarse-graining is performed by grouping eigenstates of the initial and final Hamiltonians into intervals of width $\delta \epsilon = \alpha \vert \epsilon_1 - \epsilon_0\vert$, where $\alpha$ is a coarse-graining factor that dictates the size of the width of the coarse-graining and $\{\epsilon_j\}$ are the eigenstates of the initial Hamiltonian.
The insets illustrate that coarse-grained dissipative work $\breve{W}_{\text{diss}}$ can be either smaller or larger than the fine-grained value $W_{\text{diss}}$, with $\breve{W}_{\text{diss}} \geq W_{\text{diss}}$ for $f\in [0,6]$ and $\breve{W}_{\text{diss}} < W_{\text{diss}}$ for $f>7$.}
\label{img:DissWork_RelEntropy}
\end{figure}

Fig.~\ref{img:DissWork_RelEntropy} reports the trends of fine-grained and coarse-grained dissipated work (relative entropy) and average work as the driving force varies in the range $f=[0,5]$, maintaining $\tau=1$ and $\beta = 1$. The level of coarse-graining is set by the energy resolution $\delta \epsilon = \alpha \vert \epsilon_1 - \epsilon_0 \vert$, where $\alpha$ controls the granularity of the measurement. The figure shows that the dissipated work measured on the coarse-grained states, 
$\breve{\rho}_\tau$ and $\breve{\tau}_\beta^{(\tau)}$, are higher than that measured on fine-grained states. However, starting from different initial conditions, the opposite can occur, indicating that no general rule applies. The same outcome is also observable in the case of the dissipative work.

\section{Second-law-like inequalities under coarse-graining}
\label{sec:CGSecondLaw}

Under the conditions of the coarse-grained TPM protocol, the relative entropy (dissipated work) $S(\breve{\rho}(t) \vert \vert \Theta^\dagger \breve{\tilde{\rho}}(\tau - t)\Theta)$ among the density operators in the forward and time-reversal processes in the coarse-grained TPM, $\breve{\rho}(t) = U(t,0) \breve{\tau}_\beta^{(0)} U^\dagger(t,0)$ and $\breve{\tilde{\rho}}(\tau - t) = \tilde{U}(\tau - t,0) \breve{\tau}_\beta^{(\tau)} \tilde{U}^\dagger(\tau-t,0)$ respectively, reads:
\begin{align}
\label{eqn:S1}
S(\breve{\rho}(t) \vert \vert \Theta^\dagger \breve{\tilde{\rho}}(\tau - t)\Theta) &= \mathrm{Tr}[\breve{\rho}(t) \log \breve{\rho}(t)]  \\
&~~~~-  \mathrm{Tr}[\breve{\rho}(t) \log \Theta^\dagger \breve{\tilde{\rho}}(\tau - t) \Theta]  \notag\\ 
&= \mathrm{Tr}[\breve{\tau}_\beta^{(0)} \log \breve{\tau}_\beta^{(0)}]  -  \mathrm{Tr}[\breve{\rho}(\tau) \log \breve{\tau}_{\beta}^{(\tau)}]. \notag
\end{align}
In the last equality, we used the fact that unitary evolution preserves the von Neumann entropy, along with the micro-reversibility principle for non-autonomous systems~\cite{CampisiREV}, ensuring that  $ \Theta^\dagger \tilde{U}(\tau-t, 0) \Theta = U^\dagger(\tau,t)$. Expanding the two terms in Eq.~\eqref{eqn:S1}, we obtain
\begin{subequations}
\begin{align}
\label{eqn:out1}
\mathrm{Tr}[\breve{\tau}_\beta^{(0)} \log \breve{\tau}_\beta^{(0)}] &=  - \beta \bigr(\mathrm{Tr}[\breve{H}_{0} \breve{\tau}_\beta^{(0)}] - F_{0} \bigl), \\
\mathrm{Tr}[\breve{\rho}(\tau) \log \breve{\tau}_{\beta}^{(\tau)}] &= - \beta \bigr(\mathrm{Tr}[\breve{H}_{\tau} \breve{\rho}(\tau)] - F_{\tau} \bigl).
\label{eqn:out2}
\end{align}
\end{subequations}
We note that, because the coarse-grained energies depend on the temperature, $\mathrm{Tr}[\breve{H}_{0} \breve{\tau}_\beta^{(0)}] \neq - {\partial} \, \log(\breve{Z}_{0}) / {\partial \beta}$, with $\breve{Z}_{0} = \sum_J e^{-\beta E_J^{(0)}} \mathrm{Tr}[\Pi_J]$, in contrast to the fine-grained case.

Recombining the two expressions \eqref{eqn:out1}-\eqref{eqn:out2} within Eq.~\eqref{eqn:S1}, we achieve:
\begin{align}
\label{eqn:coarse_grained_QrelEntropy}
S(\breve{\rho}(t) \vert \vert \Theta^\dagger \breve{\tilde{\rho}}(\tau - t) \Theta) = \beta \bigr(\langle{\breve{W}}\rangle - \Delta F\bigl) \geq 0,
\end{align}
where we introduced the mean `effective work' under coarse-graining $\langle{\breve{W}}\rangle = \mathrm{Tr}[\breve{H}_{\tau} \breve{\tau}_\beta^{(\tau)}]  - \mathrm{Tr}[\breve{H}_{0} \breve{\tau}_{\beta}^{(0)}]$. Eq.~\eqref{eqn:coarse_grained_QrelEntropy} establishes a fundamental link between information theory and thermodynamics at every level of the instrument's precision, linking the relative entropy between density operators in forward and time-reverse processes to the work and free energy changes of the system. Indeed, they provide a family of second-law-like relations, $\langle{\breve{W}}\rangle \geq \Delta F$, one for each possible level of coarse-graining, as directly implied by the positivity of the quantum relative entropy.

For each level of coarse-grained resolution, the quantity $\langle \breve{W}_{\text{diss}} \rangle = \langle \breve{W} \rangle - \Delta F$ can be interpreted as the average amount of work, beyond the free energy difference, required to implement the protocol $\Lambda$, as perceived by the corresponding coarse-grained agent. Importantly, because the measurement procedure itself alters the dynamics (via measurement and repreparation), 
there is, in general, no strict relationship between $\breve{W}_{\text{diss}}$ and the fine-grained dissipative work $W_{\text{diss}}$. That is, depending on the resolution, one may find either $\breve{W}_{\text{diss}} \geq W_{\text{diss}}$ or $\breve{W}_{\text{diss}} < W_{\text{diss}}$.

There is, however, a special situation in which the inequalities~\eqref{eqn:coarse_grained_QrelEntropy} (and, consequently, the family of second-law-like relations) are hierarchically ordered with respect to the level of resolution: the coarser the resolution, the smaller the relative entropy in Eq.~\eqref{eqn:coarse_grained_QrelEntropy}, and hence the smaller the amount of dissipative work detected by the coarse-graining agent. That happens when the coarse-graining operation $\mathcal{C}_{\Pi_J}$ corresponding to an agent's resolution and the unitary evolution implementing the driving protocol commute, i.e.,
\begin{equation}
\mathcal{C}_{\Pi_J^{(t)}}[U(t,0) ~\sigma~ U^\dagger(t,0)] = U(t,0) \mathcal{C}_{\Pi_J^{(0)}}(\sigma) U^\dagger(t,0),
\end{equation}
for any fine-grained density operator $\sigma$. This condition requires that the unitary evolution does not involve fine-grained degrees of freedom. In that case, we have:
\begin{align}
\beta (\langle W \rangle - \Delta F) &= S({\rho}(t) \vert \vert \Theta^\dagger \tilde{\rho}(\tau - t) \Theta) \\ &\geq  S(\mathcal{C}_{\Pi_J^{(t)}}[{\rho}(t)] \vert \vert \Theta^\dagger \mathcal{C}_{\Pi_J^{(t)}}[\tilde{\rho}(\tau - t)] \Theta) \notag \\ &= S(\breve{\rho}(t) \vert \vert \Theta^\dagger \breve{\tilde{\rho}}(\tau - t) \Theta) = \beta(\langle \breve{W} \rangle - \Delta F), \notag
\end{align}
where the inequality follows from the data processing inequality for the quantum relative entropy and the fact that $C_{\Pi_J}$ is a CPTP map.

We conclude that, when agent $C$ can fully distinguish the changes in the system's state due to the protocol $\Lambda$ (meaning those changes affect only the degrees of freedom accessible to them), the limitations in the TPM scheme's energy measurements can only lead to an underestimation of the work required for the process. On the contrary, driving protocols exceeding the resolution of the agent, may result in work payoffs either larger or smaller than $\langle W \rangle$.

\section{Fluctuation theorems under coarse-graining}
\label{sec:FTs}

We now focus on the work probability distribution in the coarse-grained TPM protocol in both forward and time-reversed processes. They are defined as:
\begin{subequations}
\begin{align} \label{eq:CGwork}
{P}(\breve{W}) = \sum_{I,J} \breve{P}_{I,J}~ \delta(\breve{W} - \breve{W}_{I,J}), \\ \label{eq:CGwork2}
\tilde{P}(\breve{W}) = \sum_{I,J} \breve{\tilde{P}}_{I,J}~ \delta(\breve{W} + \breve{W}_{I,J}),
\end{align}
\end{subequations}
with $\breve{W}_{I,J} = E_J^{(\tau)} - E_I^{(0)}$ being the work measured by the coarse-graining agent $C$ in every realization of the TPM and where $\breve{P}_{I,J}$ are defined in Eq.~\eqref{eq:P_IJ}. Micro-reversibility implies the following relation between the joint probability distributions in forward and time-reversal processes:
\begin{equation} \label{eq:detailed1}
\frac{\tilde{\breve{P}}_{I, J}}{P_{I, J}} = \left( \frac{\breve{p}_J^{(\tau)} N_I^{(0)}}{\breve{p}_I^{(0)} N_J^{(\tau)}} \right) =  e^{-\beta (\breve{W}_{I,J} - \Delta F)}.
\end{equation}
Introducing Eq.~\eqref{eq:detailed1} into the expressions \eqref{eq:CGwork}-\eqref{eq:CGwork2}, we obtain the detailed fluctuation theorem for the coarse-grained work probability distributions:
\begin{equation}\label{eq:CGdetailed}
\frac{\breve{P}(\breve{W})}{\tilde{\breve{P}}(-\breve{W})} = e^{\beta(\breve{W} - \Delta F)}.    
\end{equation}
As in the fine-grained case, the detailed theorem \eqref{eq:CGdetailed} implies the integral version:
\begin{align} \label{eq:CGintegral}
\langle e^{-\beta(\breve{W} - \Delta F)}\rangle &= \int dW \breve{P}(\breve{W}) e^{-\beta(\breve{W} - \Delta F)} \notag \\ 
&= \int dW \breve{\tilde{P}}(-\breve{W}) = 1.     
\end{align}
The above Eqs.~\eqref{eq:CGdetailed} and \eqref{eq:CGintegral} extend, respectively, Crooks work fluctuation relation~\cite{CrooksTheorem} and the Jarzynski inequality~\cite{Jarzynski:97} for the dissipative work to coarse-grained scenarios as introduced here. 

\begin{figure}[tb]
\centering
\includegraphics[width=\columnwidth]{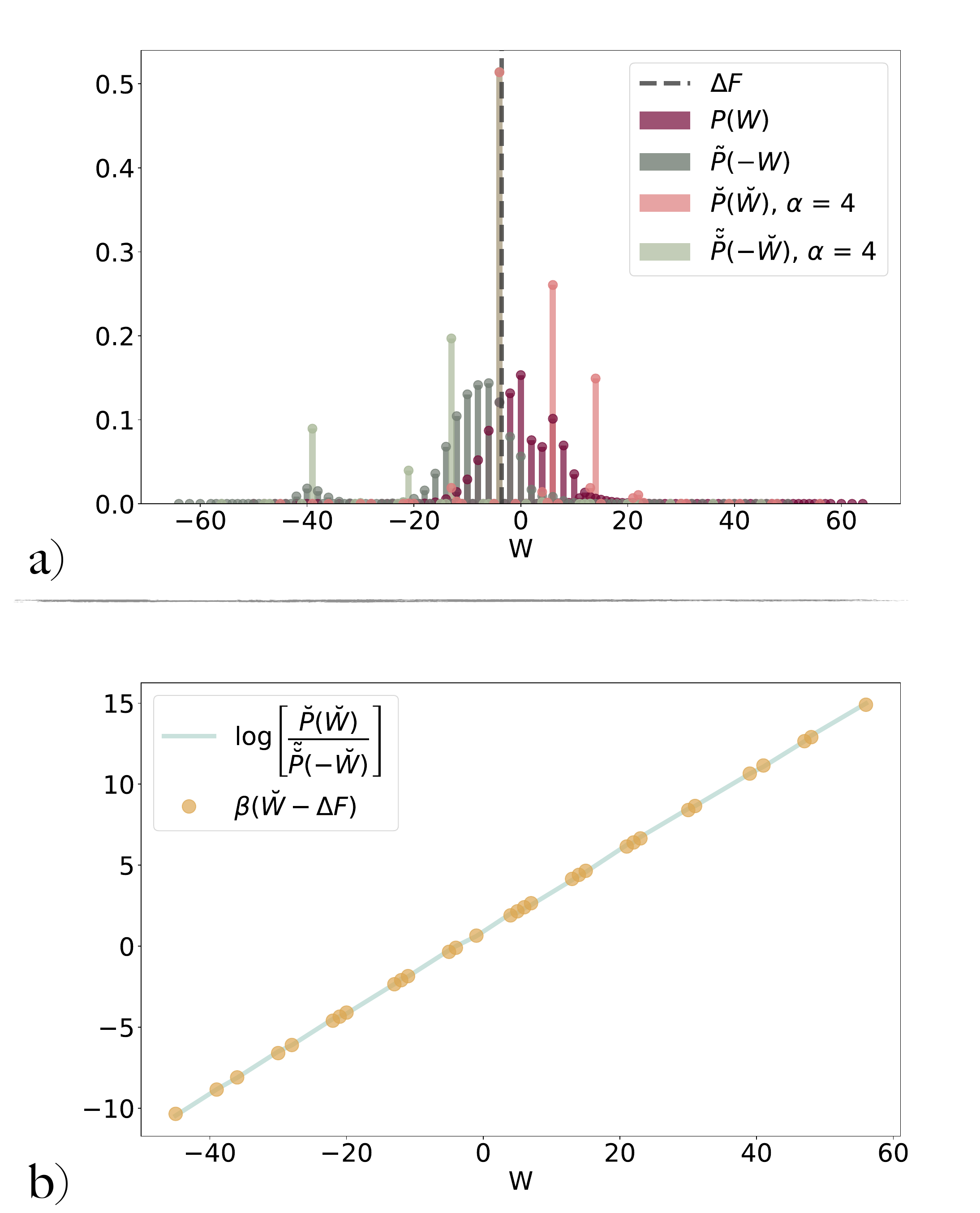}
\captionof{figure}{\textbf{\textbf{a)} Probability distributions for forward and time-reversed trajectories in fine- and coarse-grained scenarios.} Fine- and coarse-grained probability distributions for a 24-level harmonic oscillator subjected to a time-independent driving force $f = 3.8$ at inverse temperature $\beta = 1/4$. The setup is the same as in Sec.~\ref{subsec:Case_Study}, except that a 24-level system is considered here instead of 12 levels. In the fine-grained scenario, the forward and backward distributions are smooth and approximately bell-shaped, intersecting at the free energy difference $\Delta F$. In the coarse-grained case, the number of data points is significantly reduced, and the distributions appear less regular, though the intersection still occurs at the same $\Delta F$. The horizontal axis corresponds to the work $W$, expressed in units of energy. \textbf{b)} \textbf{Illustration of the fluctuation theorem for forward and time-reversed trajectories in coarse-grained scenarios.} The left- and right-hand sides of Eq.~\eqref{eq:CGdetailed} (up to a logarithm) are plotted and shown to coincide exactly, confirming that the detailed fluctuation theorem remains valid under coarse-graining.}
\label{img:GaussianFB}
\end{figure}

An important application of the detailed fluctuation theorem in Eq.~\eqref{eq:crooks} and its integral counterpart has been the determination of equilibrium free energy differences---a genuine equilibrium property---from the measurement of the work performed in (arbitrary) non-equilibrium conditions~\cite{Liphardt:02,Collin:05}. The practical relevance of Eqs.~\eqref{eq:CGdetailed} and \eqref{eq:CGintegral} is that they allow to determine the difference in equilibrium free energy from the measurements performed on the system \emph{under arbitrary resolution of the coarse-graining}, as long as $\breve{N} > 1$. Importantly, this result holds regardless of the precision with which the average dissipative work is estimated, and independently of whether the underlying unitary evolution involves fine-grained degrees of freedom. This feature is illustrated in Fig.~\ref{img:GaussianFB}. 

\section{Non-invasive coarse-graining TPM scheme}
\label{sec:Relation_to_Other_CG}

One may wish to compare the coarse-grained TPM scheme introduced in Sec.~\ref{sec:Dissipative} with a simpler variant in which the system is not reprepared following the energy measurements with coarse-grained projectors $\Pi_{I}^{(0)}$ and $\Pi_{J}^{(\tau)}$. This alternative may seem less invasive, as it avoids modifying the post-measurement state through repreparation. We refer to this protocol as the ``non-invasive TPM scheme'', which consists of the following steps: \textit{i.} the fine-grained thermal state is subjected to a coarse-grained energy measurement at $t = 0$; \textit{ii.} the \textit{projected} state $\Pi_I^{(0)} \tau_\beta^{(0)} \Pi_I^{(0)}/N_I$, corresponding to measurement outcome $I$, evolves unitarily under a unitary evolution until $t = \tau$; \textit{iii.} a second coarse-grained energy measurement is performed, yielding outcome $J$. In this case, the joint probability of obtaining outcomes $I$ and $J$ in the initial and final measurements, respectively, is given by $\bar{P}_{I,J} = \tr[\Pi_J^{(\tau)} U(\tau,0) \Pi_I^{(0)} \tau_\beta^{(0)}\Pi_I^{(0)} U^\dagger(\tau,0)]$, with an analogous expression holding for the time-reversed process.

Importantly, in the non-invasive TPM scheme, the key results derived in Secs.~\ref{sec:CGSecondLaw} and \ref{sec:FTs} (namely, the non-equilibrium relation for dissipative work in Eq.~\eqref{eqn:coarse_grained_QrelEntropy} and the fluctuation theorem in Eq.~\eqref{eq:CGdetailed}) no longer hold. Although this scheme may appear less invasive at first glance, the absence of the re-preparation step alters the structure of the process in a way that invalidates those results.
In particular, for the average dissipative work, we find:
\begin{align} \label{eq:non-invasive}
\beta (\langle W \rangle - \Delta F) &= S({\rho}(t) \vert \vert \Theta^\dagger \tilde{\rho}(\tau - t) \Theta)  \\ 
& \geq S(\bar{\rho}(t) \vert \vert \Theta^\dagger \tilde{\bar{\rho}}(\tau - t) \Theta) \neq   \beta (\langle \bar{W} \rangle - \Delta F), \notag
\end{align}
where $\bar{\rho}(t) = U(t,0) \sum_I \Pi_I^{(0)} \tau_\beta^{(0)}\Pi_I^{(0)} U^\dagger(t,0)$ is the system state at an arbitrary intermediate time $t$ of the forward process,  $\tilde{\bar{\rho}}(t) = \tilde{U}(\tau -t,0) \sum_J \Pi_J^{(\tau)} \tau_\beta^{(\tau)}\Pi_J^{(\tau)} \tilde{U}^\dagger(\tau - t,0)$ its time-reversal counterpart, and $\langle \bar{W} \rangle = \tr[\bar{\rho}(\tau) \breve{H}_\tau] - \tr[\tau_\beta^{(0)} \breve{H}_0]$ is the average work measured by the coarse-graining agent.

\begin{figure}[tb]
\centering
\includegraphics[width=\columnwidth]{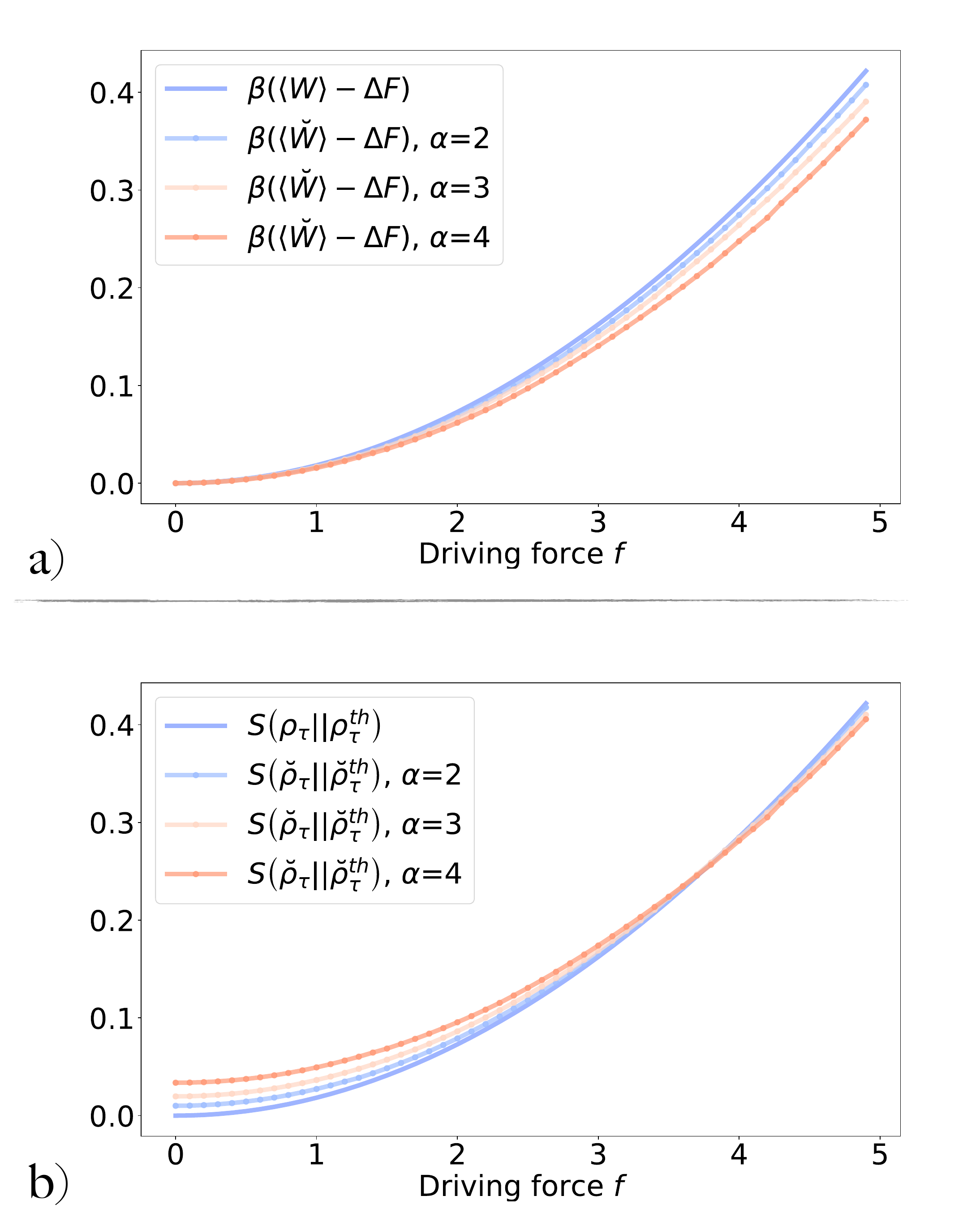}
\captionof{figure}{\textbf{Comparison between dissipative work (Panel \textbf{a)}) and relative entropy (Panel \textbf{b)}) in fine- and coarse-grained scenarios for the ``non-invasive coarse-graining'' protocol discussed in Sec.~\ref{sec:Relation_to_Other_CG}.} The figure presents the same case study as Fig.~\ref{img:DissWork_RelEntropy}, but applies the alternative ``non-invasive'' protocol, where the re-preparation step is performed using fine-grained instruments, in contrast to the standard protocol used throughout the main text. While in Fig.~\ref{img:DissWork_RelEntropy} dissipative work and relative entropy coincide (hence only one is plotted), here they differ, as seen from the mismatch between Panels \textbf{a)} and \textbf{b)}. Nonetheless, in the non-invasive case, the coarse-grained curves retain the overall shape of their fine-grained counterparts, unlike in Fig.~\ref{img:DissWork_RelEntropy}.}
\label{img:DissWork_RelEntropy_NICG}
\end{figure}

In Eq.~\eqref{eq:non-invasive}, the inequality is always warranted by means of the contractivity in the relative entropy, which ensures a useful lower bound for the estimation of the fine-grained dissipative work~\cite{Kawai2007,Parrondo2009}. However, the correspondence between the coarse-grained relative entropy and the work measured by the agent performing the energy measurements in the TPM scheme is not universal. This is due to the fact that the projected thermal states $\sum_I \Pi_I^{(0)} \tau_\beta^{(0)}\Pi_I^{(0)}$ and $\sum_J \Pi_J^{(\tau)} \tau_\beta^{(\tau)}\Pi_J^{(\tau)}$ retain the fine-grained structure of the original fine-grained states within each block and hence cannot be rewritten, in general, as thermal states with Gibbs form. Similarly, this prevents the corresponding work probability distributions $\bar{P}(\bar{W}) = \sum_{I,J}\bar{P}_{I,J}\delta(\bar{W}- \breve{W}_{I,J})$ [and the analogous expression for the time-reversed protocol $\tilde{\bar{P}}(\bar{W})$], to verify the detailed work fluctuation theorems in Eqs.~\eqref{eq:CGdetailed}-\eqref{eq:CGintegral}. This deviation from the standard scheme is illustrated in Fig.~\ref{img:DissWork_RelEntropy_NICG}, which compares dissipative work and relative entropy under the non-invasive protocol, highlighting the divergence between these two quantities despite their similar overall trends.

\section{Generalization to non-orthogonal (POVM-based) coarse graining}\label{sec:GeneralizationToPOVM}

Throughout this work, we have modelled finite measurement resolution via an orthogonal coarse-graining in the energy basis, represented by a set of projectors $\{\Pi_J\}$ forming a projective-valued measure (PVM), $\sum_J \Pi_J = \mathbb{1}$. This corresponds to a situation in which each microscopic energy eigenstate is assigned to exactly one macroscopic outcome (slot). While this captures finite resolution in the sense of limited distinguishability, it excludes scenarios in which the measurement device may also misidentify outcomes, so that a given microscopic energy may contribute to multiple macroscopic readouts. Such situations naturally arise in experiments where the reported energy is obtained by binning a continuous, noisy measurement signal, or by inferring the energy from measurements performed through an auxiliary degree of freedom (e.g., via population measurements or fluorescence signals)~\cite{Itano1993, Keselman_2011, Zhang17}, so that the same microscopic energy can lead to different macroscopic readouts with nonzero probability.

A consistent mathematical description of overlapping coarse-graining is provided by replacing the PVM by a positive-operator valued measure (POVM) $\{F_J\}$, where each outcome $J$ is associated with a positive operator $F_J \geq 0$ satisfying $\sum_J F_J = \mathbb{1}$.
In the present thermodynamic setting, it is natural to restrict to POVMs that are diagonal in the energy eigenbasis $\{\ket{\epsilon_j}\}$, so that
\begin{equation}
    F_J = \sum_j w_{J|j} \ket{\epsilon_j}\bra{\epsilon_j}, \label{eq:POVM_energy}
\end{equation}
with $w_{J|j} \in [0,1]$, $\sum_J w_{J|j} = 1 \, \forall j$. 
The coefficients $w_{J|j}$ admit a direct operational interpretation as conditional probabilities: $w_{J|j}$ is the probability that the measurement apparatus reports outcome $J$ when the system is in the energy eigenstate $\ket{\epsilon_j}$. Overlapping coarse-graining corresponds to the case in which, for a fixed $j$, more than one coefficient $w_{J|j}$ is nonzero.
A physically motivated choice of such weights can be constructed by introducing nominal energy values $\{E_J\}$ associated with the measurement outcomes and a response function $g$ describing the finite resolution of the detector, for instance
\begin{equation}
    w_{J|j} = \frac{e^{-(\epsilon_j - E_J)^2/(2\sigma^2)}}
{\sum_{K} e^{-(\epsilon_j - E_K)^2/(2\sigma^2)}},
\end{equation}
where $\sigma$ quantifies the energy uncertainty of the readout. In the limit $\sigma \to 0$, the POVM effects approach orthogonal projectors onto sharp energy intervals, recovering the PVM coarse-graining employed in previous sections.

Given a POVM $\{F_J\}$, the natural analogue of the coarse-graining operation introduced in Eq.~\eqref{eq:cgmap} is a measure-and-
reprepare channel of the form
\begin{equation} \label{eq:POVM_channel}
    \mathcal{C}_F(\rho) = \sum_J \tr\bigl[F_J \rho\bigr] \, \sigma_J,
\end{equation}
where $\sigma_J$ is the post-measurement state associated with outcome $J$. A particularly natural and unbiased choice (which generalizes the flat re-preparation within each slot used in the orthogonal case) is $\sigma_J = {F_J}/{\tr[F_J]}$.
With this choice, the channel is completely positive, trace preserving, and unital, since
\begin{equation}
    \mathcal{C}_F(\mathbb{1}) = \sum_J \tr[F_J]\frac{F_J}{\tr[F_J]} = \sum_J F_J = \mathbb{1}.
\end{equation}
In the special case $F_J = \Pi_J$, this construction reduces exactly to the coarse-graining map $\mathcal{C}_{\Pi_J}$ used throughout the paper.
The observable effectively measured by the coarse-grained apparatus can be defined as
\begin{equation} \label{eq:POVM_energy_operator}
    \breve{H} = \sum_J E_J F_J,
\end{equation}
which, for diagonal POVMs, takes the form $\breve{H} = \sum_j \breve{\epsilon}_j \ket{\epsilon_j}\bra{\epsilon_j}$, with $\breve{\epsilon}_j := \sum_J E_J w_{J|j}$.
Hence, overlapping coarse-graining induces a smooth deformation of the microscopic energy spectrum into effective energies $\breve{\epsilon}_j$ reflecting both limited resolution and readout errors. Expectation values of the measured energy are then given by $\langle \breve{H} \rangle_\rho = \tr[\rho \breve{H}]$.

While this POVM-based framework provides a consistent and experimentally motivated model of overlapping coarse graining, it leads to qualitative differences with respect to the orthogonal case considered in this work. In particular, POVM outcomes do not define a decomposition of Hilbert space into disjoint energy sectors, and the associated coarse-graining channel cannot be interpreted as flattening independent subspaces. As a consequence, repeated coarse graining generally continues to modify the state, and the channel $\mathcal{C}_F$ is not idempotent, $\mathcal{C}_F^2 \neq \mathcal{C}_F$.

From a thermodynamic perspective, however, the more crucial structural property is unitality. For the choice $\sigma_J = F_J/\mathrm{Tr}[F_J]$, the coarse-graining map $\mathcal{C}_F$ remains unital, which is the key condition ensuring the validity of quantum Jarzynski-type equalities for general quantum channels~\cite{Rastegin2013}. In contrast, if the coarse-graining instrument is not unital, additional correction terms appear in fluctuation relations~\cite{Albash2013,Rastegin2014}, reflecting entropy production associated with the measurement back-action itself~\cite{Manzano2015}.

Despite unitality, the lack of idempotence and of a sharp sector structure has important consequences. In particular, coarse-grained thermal states $\mathcal{C}_F(\tau_\beta)$ do not generally retain an exact Gibbs form with respect to the effective Hamiltonian $\breve{H}$, and the forward and backward coarse-grained dynamics are no longer related by a simple time-reversal symmetry. As a result, similarly to the case of TPMs with non-ideal measurements~\cite{Debarba:19,Ito:19}, Crooks-type fluctuation relations are not generically satisfied in the presence of non-orthogonal (POVM-based) coarse graining.

Consequently, while Jarzynski relations may still be recovered for suitably chosen unital POVM-based coarse-graining maps, the invariance of the full set of thermodynamic relations derived in this work (most notably Crooks relations) relies on the stronger structure provided by orthogonal, projective coarse graining. Investigating the precise conditions under which Crooks-type relations and second-law-like inequalities persist for non-orthogonal coarse graining constitutes an interesting direction for future research.

\section*{DISCUSSION}
\label{sec:Discussion}

In this work, we have explored the thermodynamic implications of coarse-graining in the energy basis from an operational point of view. We have seen that coarse-graining as a quantum operation has energetic implications for information thermodynamics, and that, consequently, to be thermodynamically consistent, coarse-grained energies must be selected in a very precise way. Such a choice of energies leads to the definition of coarse-graining thermal states, which are the basis for obtaining several known non-equilibrium relations by only using the information retrieved by an agent who does not dispose of measuring instruments of infinite precision. This focus on energy-diagonal coarse graining reflects a physical modelling choice tailored to finite energy resolution. More general forms of coarse graining, including restrictions on other degrees of freedom or coarse-grained observables not commuting with the Hamiltonian, may be described within a similar mathematical framework but can lead to qualitatively different thermodynamic behaviour (a discussion of these cases and their limitations is provided in Appendix~\ref{app:D}).

We found that the fundamental thermodynamic relations remain formally identical to those obtained using instruments of infinite precision. However, the physical conclusions drawn from coarse-grained quantities can differ significantly from those based on their fine-grained counterparts. For example, when the underlying dynamics involve degrees of freedom that are inaccessible to the coarse-graining agent, the work consumed or produced during a driving protocol may deviate substantially from the value obtained using fine-grained instruments. Importantly, our assessment of work in this study focuses on the system dynamics and does not include additional thermodynamic costs associated with the implementation of the TPM scheme itself. Such costs (arising, for instance, from the realization of projective measurements~\cite{Guryanova2020,Taranto2023}) can be incorporated by extending the TPM framework to account for non-ideal measurement processes~\cite{Debarba:19}.

We illustrated our findings using a 12-level harmonic oscillator subjected to a thermodynamic evolution driven by a time-independent external force. While in the ideal (fine-grained) case the evolution neither requires nor produces work, we observed that increasing the driving strength leads to a growing amount of effective work production from the perspective of a coarse-grained agent. We also examined how the work probability distributions associated with the forward and time-reversed protocols change with the level of coarse-graining, and showed that equilibrium free energies can still be accurately estimated from coarse-grained measurements of non-equilibrium work, regardless of the measurement precision.  

In conclusion, our results establish a consistent thermodynamic framework for scenarios in which only partial information (limited by the finite energy resolution of experimental instruments) is available. We examined the impact of limited-precision measurements on the estimation of key quantum thermodynamic quantities and showed that, although the individual quantities depend on the measurement resolution, the fundamental relations among them remain agent-independent. These findings open new perspectives in the investigation of the theory of quantum measurements in quantum thermodynamics and, in particular, in its relevance in any experimental scheme seeking to address the concepts of reversibility and the quantum thermodynamic time's arrow.

\vspace*{3 mm}

\section*{Acknowledgements} We thank M. Horodecki, G. Landi and L. Tesser for useful discussions.  \textbf{Funding:} G.R.~acknowledges financial support from the Royal Commission for the Exhibition of 1851 through a Research Fellowship, from the European Commission through Advanced Grant ERC-2020-ADG101021085 (FLQuant), and from EPSRC through Standard Proposal Grant EP/X016218/1 (Mono-Squeeze). \v{C}.B.~acknowledges financial support from the Austrian Science Fund (FWF) No. 10.55776/F71
and 10.55776/COE1, and the John
Templeton Foundation (grant IDs 61466 and 62312) as part of the ``Quantum Information Structure of Spacetime (QISS)'' project (qiss.fr). G.M.~acknowledges financial support from the Ram\'on y Cajal program RYC2021-031121-I funded by MICIU/AEI/10.13039/501100011033 and European Union NextGenerationEU/PRTR, the CoQuSy project PID2022-140506NB-C21 and C22, and the María de Maeztu project CEX2021-001164-M for Units of Excellence, funded by MICIU/AEI/10.13039/501100011033/FEDER, UE. For the purpose of open access, the author(s) have applied a CC-BY license to any author-accepted manuscript version arising from this submission.
\textbf{Data Availability:} All codes used to produce the data are available at~\cite{RubinoCGQThermoCode}.

\filbreak
\renewcommand{\baselinestretch}{1.2} 
\bibliography{References}

@article{Aberg2018,
  title = {Fully Quantum Fluctuation Theorems},
  author = {\AA{}berg, Johan},
  journal = {Phys. Rev. X},
  volume = {8},
  issue = {1},
  pages = {011019},
  numpages = {78},
  year = {2018},
  month = {Feb},
  publisher = {American Physical Society},
  doi = {10.1103/PhysRevX.8.011019},
  url = {https://link.aps.org/doi/10.1103/PhysRevX.8.011019}
}

@article{Manzano18b,
  title = {Quantum Fluctuation Theorems for Arbitrary Environments: Adiabatic and Nonadiabatic Entropy Production},
  author = {Manzano, Gonzalo and Horowitz, Jordan M. and Parrondo, Juan M. R.},
  journal = {Phys. Rev. X},
  volume = {8},
  issue = {3},
  pages = {031037},
  numpages = {26},
  year = {2018},
  month = {Aug},
  publisher = {American Physical Society},
  doi = {10.1103/PhysRevX.8.031037},
  url = {https://link.aps.org/doi/10.1103/PhysRevX.8.031037}
}

@article{Guryanova2020,
  doi = {10.22331/q-2020-01-13-222},
  url = {https://doi.org/10.22331/q-2020-01-13-222},
  title = {Ideal {P}rojective {M}easurements {H}ave {I}nfinite {R}esource {C}osts},
  author = {Guryanova, Yelena and Friis, Nicolai and Huber, Marcus},
  journal = {{Quantum}},
  issn = {2521-327X},
  publisher = {{Verein zur F{\"{o}}rderung des Open Access Publizierens in den Quantenwissenschaften}},
  volume = {4},
  pages = {222},
  month = jan,
  year = {2020}
}

@article{Taranto2023,
  title = {Landauer Versus Nernst: What is the True Cost of Cooling a Quantum System?},
  author = {Taranto, Philip and Bakhshinezhad, Faraj and Bluhm, Andreas and Silva, Ralph and Friis, Nicolai and Lock, Maximilian P.E. and Vitagliano, Giuseppe and Binder, Felix C. and Debarba, Tiago and Schwarzhans, Emanuel and Clivaz, Fabien and Huber, Marcus},
  journal = {PRX Quantum},
  volume = {4},
  issue = {1},
  pages = {010332},
  numpages = {61},
  year = {2023},
  month = {Mar},
  publisher = {American Physical Society},
  doi = {10.1103/PRXQuantum.4.010332},
  url = {https://link.aps.org/doi/10.1103/PRXQuantum.4.010332}
}

@article{Rastegin2013,
doi = {10.1088/1742-5468/2013/06/P06016},
url = {https://doi.org/10.1088/1742-5468/2013/06/P06016},
year = {2013},
month = {jun},
publisher = {IOP Publishing and SISSA},
volume = {2013},
number = {06},
pages = {P06016},
author = {Rastegin, Alexey E},
title = {Non-equilibrium equalities with unital quantum channels},
journal = {Journal of Statistical Mechanics: Theory and Experiment}
}

@article{Albash2013,
  title = {Fluctuation theorems for quantum processes},
  author = {Albash, Tameem and Lidar, Daniel A. and Marvian, Milad and Zanardi, Paolo},
  journal = {Phys. Rev. E},
  volume = {88},
  issue = {3},
  pages = {032146},
  numpages = {14},
  year = {2013},
  month = {Sep},
  publisher = {American Physical Society},
  doi = {10.1103/PhysRevE.88.032146},
  url = {https://link.aps.org/doi/10.1103/PhysRevE.88.032146}
}

@article{Rastegin2014,
  title = {Jarzynski equality for quantum stochastic maps},
  author = {Rastegin, Alexey E. and \ifmmode \dot{Z}\else \.{Z}\fi{}yczkowski, Karol},
  journal = {Phys. Rev. E},
  volume = {89},
  issue = {1},
  pages = {012127},
  numpages = {10},
  year = {2014},
  month = {Jan},
  publisher = {American Physical Society},
  doi = {10.1103/PhysRevE.89.012127},
  url = {https://link.aps.org/doi/10.1103/PhysRevE.89.012127}
}

@article{Manzano2015,
  title = {Nonequilibrium potential and fluctuation theorems for quantum maps},
  author = {Manzano, Gonzalo and Horowitz, Jordan M. and Parrondo, Juan M. R.},
  journal = {Phys. Rev. E},
  volume = {92},
  issue = {3},
  pages = {032129},
  numpages = {9},
  year = {2015},
  month = {Sep},
  publisher = {American Physical Society},
  doi = {10.1103/PhysRevE.92.032129},
  url = {https://link.aps.org/doi/10.1103/PhysRevE.92.032129}
}

@book{Callen_book,
title = {Thermodynamics and an introduction to Thermostatistics},
author = {Callen, Herbert B.},
year = {1985},
publisher = {Jonh Willey and Sons},
note = {Second Ed.}
}

@article{Jaynes1957a,
  title = {Information Theory and Statistical Mechanics},
  author = {Jaynes, E. T.},
  journal = {Phys. Rev.},
  volume = {106},
  issue = {4},
  pages = {620--630},
  numpages = {0},
  year = {1957},
  month = {May},
  publisher = {American Physical Society},
  doi = {10.1103/PhysRev.106.620},
  url = {https://link.aps.org/doi/10.1103/PhysRev.106.620}
}

@article{Jaynes1957b,
  title = {Information Theory and Statistical Mechanics. II},
  author = {Jaynes, E. T.},
  journal = {Phys. Rev.},
  volume = {108},
  issue = {2},
  pages = {171--190},
  numpages = {0},
  year = {1957},
  month = {Oct},
  publisher = {American Physical Society},
  doi = {10.1103/PhysRev.108.171},
  url = {https://link.aps.org/doi/10.1103/PhysRev.108.171}
}

@article{Szilard1929,
  title = {über die Entropieverminderung in einem thermodynamischen System bei Eingriffen intelligenter Wesen},
  author = {Szilard, L. },
  journal = {Zeitschrift für Physik},
  volume = {53},
  issue = {11},
  pages = {840--856},
  year = {1929},
  month = {Nov},
  doi = {10.1007/BF01341281}
}

@article{Landauer1961,
  author={R. {Landauer}},
  journal={IBM Journal of Research and Development}, 
  title={Irreversibility and Heat Generation in the Computing Process}, 
  year={1961},
  volume={5},
  number={3},
  pages={183-191},
  }

@article{Bennett1982,
  title = {The thermodynamics of computation—a review},
  author = {Bennett, Charles H. },
  journal = {International Journal of Theoretical Physics},
  volume = {21},
  issue = {12},
  pages = {905--940},
  year = {1982},
  month = {Dec},
  doi = {10.1007/BF02084158}
}

@article{CampisiREV,
  title = {Colloquium: Quantum fluctuation relations: Foundations and applications},
  author = {Campisi, Michele and H\"anggi, Peter and Talkner, Peter},
  journal = {Rev. Mod. Phys.},
  volume = {83},
  issue = {3},
  pages = {771--791},
  numpages = {0},
  year = {2011},
  month = {Jul},
  publisher = {American Physical Society},
  doi = {10.1103/RevModPhys.83.771},
  url = {https://link.aps.org/doi/10.1103/RevModPhys.83.771}
}

@article{CrooksTheorem,
  title = {Entropy production fluctuation theorem and the nonequilibrium work relation for free energy differences},
  author = {Crooks, Gavin E.},
  journal = {Phys. Rev. E},
  volume = {60},
  issue = {3},
  pages = {2721--2726},
  numpages = {0},
  year = {1999},
  month = {Sep},
  publisher = {American Physical Society},
  doi = {10.1103/PhysRevE.60.2721},
  url = {https://link.aps.org/doi/10.1103/PhysRevE.60.2721}
}

@book{MaxwellDemon,
 URL = {http://www.jstor.org/stable/j.ctt7zts1p},
  publisher = {Princeton University Press},
 title = {Maxwell's Demon 2: Entropy, Information, Computing},
 year = {2002},
author = {Leff, Harvey S. and Rex, Andrew F.}
}

@article{Goold_2016,
	doi = {10.1088/1751-8113/49/14/143001},
	year = 2016,
	month = {feb},
	journal = {Journal of Physics A: Mathematical and Theoretical},
	publisher = {{IOP} Publishing},
	volume = {49},
	number = {14},
	pages = {143001},
	author = {John Goold and Marcus Huber and Arnau Riera and L{\'{\i}}dia del Rio and Paul Skrzypczyk},
	title = {The role of quantum information in thermodynamics{\textemdash}a topical review}
}

@Article{Strasberg24,
	title={{Comparative microscopic study of entropies and their production}},
	author={Philipp Strasberg and Joseph Schindler},
	journal={SciPost Phys.},
	volume={17},
	pages={143},
	year={2024},
	publisher={SciPost},
	doi={10.21468/SciPostPhys.17.5.143},
	url={https://scipost.org/10.21468/SciPostPhys.17.5.143},
}

@article{Kawai2007,
  title = {Dissipation: The Phase-Space Perspective},
  author = {Kawai, R. and Parrondo, J. M. R. and den Broeck, C. Van},
  journal = {Phys. Rev. Lett.},
  volume = {98},
  issue = {8},
  pages = {080602},
  numpages = {4},
  year = {2007},
  month = {Feb},
  publisher = {American Physical Society},
  doi = {10.1103/PhysRevLett.98.080602},
  url = {https://link.aps.org/doi/10.1103/PhysRevLett.98.080602}
}

@article{Parrondo2009,
	doi = {10.1088/1367-2630/11/7/073008},
	year = 2009,
	month = {jul},
	publisher = {{IOP} Publishing},
	volume = {11},
	number = {7},
	pages = {073008},
	author = {J. M. R. Parrondo and C. Van den Broeck and R. Kawai},
	title = {Entropy production and the arrow of time},
	journal = {New Journal of Physics},
}

@inbook{Sagawa,
author = {Sagawa, Takahiro},
title = {Second Law-Like Inequalities with Quantum Relative Entropy: An Introduction},
booktitle = {Lectures on Quantum Computing, Thermodynamics and Statistical Physics},
chapter = {},
publisher = {World Scientific Publishing},
year = {2012},
pages = {125-190},
doi = {10.1142/9789814425193_0003},
URL = {https://www.worldscientific.com/doi/abs/10.1142/9789814425193_0003}
}

@article{Guryanova2016,
	doi = {10.1038/ncomms12049},
	year = {2016},
	month = {Jul},
	publisher = {},
	volume = {7},
	number = {1},
	pages = {12049},
	author = {Guryanova, Yelena and Popescu, Sandu and Short, Anthony J. and Silva, Ralph and Skrzypczyk, Paul},
	title = {Thermodynamics of quantum systems with multiple conserved quantities},
	journal = {Nature Communications},
}

@article{YungerHalpern2016a,
archivePrefix = {arXiv},
arxivId = {1512.01189},
author = {{Yunger Halpern}, Nicole and Faist, Philippe and Oppenheim, Jonathan and Winter, Andreas},
doi = {10.1038/ncomms12051},
eprint = {1512.01189},
isbn = {9781137332875},
issn = {20411723},
journal = {Nature Communications},
number = {May},
pages = {1--7},
pmid = {27384494},
publisher = {Nature Publishing Group},
title = {{Microcanonical and resource-theoretic derivations of the thermal state of a quantum system with noncommuting charges}},
url = {http://dx.doi.org/10.1038/ncomms12051},
volume = {7},
year = {2016}
}

@article{Manzano2018,
  title = {Optimal Work Extraction and Thermodynamics of Quantum Measurements and Correlations},
  author = {Manzano, Gonzalo and Plastina, Francesco and Zambrini, Roberta},
  journal = {Phys. Rev. Lett.},
  volume = {121},
  issue = {12},
  pages = {120602},
  numpages = {7},
  year = {2018},
  month = {Sep},
  publisher = {American Physical Society},
  doi = {10.1103/PhysRevLett.121.120602},
  url = {https://link.aps.org/doi/10.1103/PhysRevLett.121.120602}
}

@article{Parrondo2015,
  title                    = {Thermodynamics of information},
  author                   = {Parrondo, J. M. R. and Horowitz, J. M. and Sagawa, T.},
  journal                  = {Nat. Phys.},
  year                     = {2015},
  month                    = feb,
  pages                    = {131},
  volume                   = {11},
  doi                      = {10.1038/nphys3230},
  publisher                = {Nature Publishing Group},
}

@article{Popescu2014,
  title                    = {Work extraction and thermodynamics for individual quantum systems},
  author                   = {Skrzypczyk, P. and Short, A. J. and Popescu, S.},
  journal                  = {Nat. Commun.},
  year                     = {2014},
  pages                    = {4185},
  volume                   = {5},
  publisher                = {Nature Publishing Group, a division of Macmillan Publishers Limited. All Rights Reserved.},
  url                      = {http://dx.doi.org/10.1038/ncomms5185}
}

@article{Anders2016,
  title                    = {Coherence and measurement in quantum thermodynamics},
  author                   = {Kammerlander, P. and Anders, J. },
  journal                  = {Sci. Rep.},
  year                     = {2016},
  month                    = {Feb},
  pages                    = {22174},
  volume                   = {6},
  doi                      = {10.1038/srep22174},
  url                      = {https://www.nature.com/articles/srep22174}
}

@article{VedralRev,
  title = {The role of relative entropy in quantum information theory},
  author = {Vedral, V.},
  journal = {Rev. Mod. Phys.},
  volume = {74},
  issue = {1},
  pages = {197--234},
  numpages = {0},
  year = {2002},
  month = {Mar},
  publisher = {American Physical Society},
  doi = {10.1103/RevModPhys.74.197},
  url = {https://link.aps.org/doi/10.1103/RevModPhys.74.197}
}

@article{Esposito2011,
	doi = {10.1209/0295-5075/95/40004},
	url = {https://doi.org/10.1209%2F0295-5075%2F95%2F40004},
	year = 2011,
	month = {aug},
	publisher = {{IOP} Publishing},
	volume = {95},
	number = {4},
	pages = {40004},
	author = {M. Esposito and C. Van den Broeck},
	title = {Second law and Landauer principle far from equilibrium},
	journal = {{EPL} (Europhysics Letters)}
}

@article {Liphardt:02,
	author = {Liphardt, Jan and Dumont, Sophie and Smith, Steven B. and Tinoco, Ignacio and Bustamante, Carlos},
	title = {Equilibrium Information from Nonequilibrium Measurements in an Experimental Test of Jarzynski{\textquoteright}s Equality},
	volume = {296},
	number = {5574},
	pages = {1832--1835},
	year = {2002},
	doi = {10.1126/science.1071152},
	publisher = {American Association for the Advancement of Science},
	issn = {0036-8075},
	URL = {https://science.sciencemag.org/content/296/5574/1832},
	eprint = {https://science.sciencemag.org/content/296/5574/1832.full.pdf},
	journal = {Science}
}

@article{Collin:05,
       author = {{Collin}, D. and {Ritort}, F. and {Jarzynski}, C. and {Smith}, S.~B. and {Tinoco}, I. and {Bustamante}, C.},
        title = {Verification of the Crooks fluctuation theorem and recovery of RNA folding free energies},
      journal = {Nature},
         year = 2005,
        month = sep,
       volume = {437},
       number = {7056},
        pages = {231-234},
          doi = {10.1038/nature04061}
}

@article{Esposito:12,
  title = {Stochastic thermodynamics under coarse graining},
  author = {Esposito, Massimiliano},
  journal = {Phys. Rev. E},
  volume = {85},
  issue = {4},
  pages = {041125},
  numpages = {11},
  year = {2012},
  month = {Apr},
  publisher = {American Physical Society},
  doi = {10.1103/PhysRevE.85.041125},
  url = {https://link.aps.org/doi/10.1103/PhysRevE.85.041125}
}

@article{Celani:17,
title = "Multiple-scale stochastic processes: Decimation, averaging and beyond",
journal = "Physics Reports",
volume = "670",
pages = "1 - 59",
year = "2017",
note = "Multiple-scale stochastic processes: decimation, averaging and beyond",
issn = "0370-1573",
doi = "https://doi.org/10.1016/j.physrep.2016.12.003",
url = "http://www.sciencedirect.com/science/article/pii/S0370157316303945",
author = "Stefano Bo and Antonio Celani"
}

@article{Safranek:19,
  title = {Quantum coarse-grained entropy and thermodynamics},
  author = {\ifmmode \check{S}\else \v{S}\fi{}afr\'anek, Dominik and Deutsch, J. M. and Aguirre, Anthony},
  journal = {Phys. Rev. A},
  volume = {99},
  issue = {1},
  pages = {010101},
  numpages = {6},
  year = {2019},
  month = {Jan},
  publisher = {American Physical Society},
  doi = {10.1103/PhysRevA.99.010101},
  url = {https://link.aps.org/doi/10.1103/PhysRevA.99.010101}
}

@article{Safranek2:19,
  title = {Quantum coarse-grained entropy and thermalization in closed systems},
  author = {\ifmmode \check{S}\else \v{S}\fi{}afr\'anek, Dominik and Deutsch, J. M. and Aguirre, Anthony},
  journal = {Phys. Rev. A},
  volume = {99},
  issue = {1},
  pages = {012103},
  numpages = {38},
  year = {2019},
  month = {Jan},
  publisher = {American Physical Society},
  doi = {10.1103/PhysRevA.99.012103},
  url = {https://link.aps.org/doi/10.1103/PhysRevA.99.012103}
}

@article{Nagasawa:24,
  title = {Generic increase of observational entropy in isolated systems},
  author = {Nagasawa, Teruaki and Kato, Kohtaro and Wakakuwa, Eyuri and Buscemi, Francesco},
  journal = {Phys. Rev. Res.},
  volume = {6},
  issue = {4},
  pages = {043327},
  numpages = {10},
  year = {2024},
  month = {Dec},
  publisher = {American Physical Society},
  doi = {10.1103/PhysRevResearch.6.043327},
  url = {https://link.aps.org/doi/10.1103/PhysRevResearch.6.043327}
}

@misc{Schindler25,
      title={Unification of observational entropy with maximum entropy principles}, 
      author={Joseph Schindler and Philipp Strasberg and Niklas Galke and Andreas Winter and Michael G. Jabbour},
      year={2025},
      eprint={2503.15612},
      archivePrefix={arXiv},
      primaryClass={quant-ph},
      url={https://arxiv.org/abs/2503.15612}, 
}

@article{Seifert19,
   author = "Seifert, Udo",
   title = "From Stochastic Thermodynamics to Thermodynamic Inference", 
   journal= "Annual Review of Condensed Matter Physics",
   year = "2019",
   volume = "10",
   number = "Volume 10, 2019",
   pages = "171-192",
   doi = "https://doi.org/10.1146/annurev-conmatphys-031218-013554",
   url = "https://www.annualreviews.org/content/journals/10.1146/annurev-conmatphys-031218-013554",
   publisher = "Annual Reviews",
   issn = "1947-5462",
   type = "Journal Article"
  }

@article{Vulpiani:10,
  title={Entropy production and coarse graining in Markov processes},
  author={A. Puglisi and S. Pigolotti and L. Rondoni and A. Vulpiani},
  journal={Journal of Statistical Mechanics: Theory and Experiment},
  year={2010},
  volume={2010},
  pages={05015}
}

@article{Rahav:07,
	doi = {10.1088/1742-5468/2007/09/p09012},
	url = {https://doi.org/10.1088/1742-5468/2007/09/p09012},
	year = 2007,
	month = {sep},
	publisher = {{IOP} Publishing},
	volume = {2007},
	number = {09},
	pages = {P09012--P09012},
	author = {Saar Rahav and Christopher Jarzynski},
	title = {Fluctuation relations and coarse-graining},
	journal = {Journal of Statistical Mechanics: Theory and Experiment}
}

@article{GomezMarin:08,
  title = {Lower bounds on dissipation upon coarse graining},
  author = {Gomez-Marin, A. and Parrondo, J. M. R. and Van den Broeck, C.},
  journal = {Phys. Rev. E},
  volume = {78},
  issue = {1},
  pages = {011107},
  numpages = {11},
  year = {2008},
  month = {Jul},
  publisher = {American Physical Society},
  doi = {10.1103/PhysRevE.78.011107},
  url = {https://link.aps.org/doi/10.1103/PhysRevE.78.011107}
}

@article{Esposito:15,
  title = {Stochastic thermodynamics of hidden pumps},
  author = {Esposito, Massimiliano and Parrondo, Juan M. R.},
  journal = {Phys. Rev. E},
  volume = {91},
  issue = {5},
  pages = {052114},
  numpages = {7},
  year = {2015},
  month = {May},
  publisher = {American Physical Society},
  doi = {10.1103/PhysRevE.91.052114},
  url = {https://link.aps.org/doi/10.1103/PhysRevE.91.052114}
}

@article{Shiriashi:15,
  title = {Fluctuation theorem for partially masked nonequilibrium dynamics},
  author = {Shiraishi, Naoto and Sagawa, Takahiro},
  journal = {Phys. Rev. E},
  volume = {91},
  issue = {1},
  pages = {012130},
  numpages = {7},
  year = {2015},
  month = {Jan},
  publisher = {American Physical Society},
  doi = {10.1103/PhysRevE.91.012130},
  url = {https://link.aps.org/doi/10.1103/PhysRevE.91.012130}
}

@article{Vollmer:12,
  title = {Fluctuation-Preserving Coarse Graining for Biochemical Systems},
  author = {Altaner, Bernhard and Vollmer, J\"urgen},
  journal = {Phys. Rev. Lett.},
  volume = {108},
  issue = {22},
  pages = {228101},
  numpages = {5},
  year = {2012},
  month = {May},
  publisher = {American Physical Society},
  doi = {10.1103/PhysRevLett.108.228101},
  url = {https://link.aps.org/doi/10.1103/PhysRevLett.108.228101}
}

@article{VonNeumann:10,
  title = {Proof of the Ergodic Theorem and the H-Theorem in Quantum Mechanics},
  author = {von Neumann, John},
  journal = {European Phys. J. H},
  volume = {35},
  pages = {201-237},
  year = {2010},
  doi = {10.1140/epjh/e2010-00008-5},
  note = {German original in Zeitschrift fuer Physik 57: 30-70 (1929)}
}

@article{Busch:93,
       author = {{Busch}, Paul and {Quadt}, Ralf},
        title = "{Concepts of coarse graining in quantum mechanics}",
      journal = {International Journal of Theoretical Physics},
         year = 1993,
        month = dec,
       volume = {32},
       number = {12},
        pages = {2261-2269},
          doi = {10.1007/BF00672998},
}

@article{Debarba:19,
	doi = {10.1088/1367-2630/ab4d9d},
	url = {https://doi.org/10.1088/1367-2630/ab4d9d},
	year = 2019,
	month = {nov},
	publisher = {{IOP} Publishing},
	volume = {21},
	number = {11},
	pages = {113002},
	author = {Tiago Debarba and Gonzalo Manzano and Yelena Guryanova and Marcus Huber and Nicolai Friis},
	title = {Work estimation and work fluctuations in the presence of non-ideal measurements},
	journal = {New Journal of Physics}
}

@article{Anders:13,
	doi = {10.1088/1367-2630/15/3/033022},
	url = {https://doi.org/10.1088/1367-2630/15/3/033022},
	year = 2013,
	month = {mar},
	publisher = {{IOP} Publishing},
	volume = {15},
	number = {3},
	pages = {033022},
	author = {Janet Anders and Vittorio Giovannetti},
	title = {Thermodynamics of discrete quantum processes},
	journal = {New Journal of Physics}
}

@article{Takara:10,
title = "Generalization of the second law for a transition between nonequilibrium states",
journal = "Physics Letters A",
volume = "375",
number = "2",
pages = "88 - 92",
year = "2010",
issn = "0375-9601",
doi = "https://doi.org/10.1016/j.physleta.2010.11.002",
url = "http://www.sciencedirect.com/science/article/pii/S0375960110014507",
author = "K. Takara and H.-H. Hasegawa and D.J. Driebe"
}

@article{Mayurama:09,
  title = {Colloquium: The physics of Maxwell's demon and information},
  author = {Maruyama, Koji and Nori, Franco and Vedral, Vlatko},
  journal = {Rev. Mod. Phys.},
  volume = {81},
  issue = {1},
  pages = {1--23},
  numpages = {0},
  year = {2009},
  month = {Jan},
  publisher = {American Physical Society},
  doi = {10.1103/RevModPhys.81.1},
  url = {https://link.aps.org/doi/10.1103/RevModPhys.81.1}
}

@article{Ito:19,
  title = {Generalized energy measurements and quantum work compatible with fluctuation theorems},
  author = {Ito, Kosuke and Talkner, Peter and Venkatesh, B. Prasanna and Watanabe, Gentaro},
  journal = {Phys. Rev. A},
  volume = {99},
  issue = {3},
  pages = {032117},
  numpages = {11},
  year = {2019},
  month = {Mar},
  publisher = {American Physical Society},
  doi = {10.1103/PhysRevA.99.032117},
  url = {https://link.aps.org/doi/10.1103/PhysRevA.99.032117}
}

@article{Polettini:17,
  title = {Effective Thermodynamics for a Marginal Observer},
  author = {Polettini, Matteo and Esposito, Massimiliano},
  journal = {Phys. Rev. Lett.},
  volume = {119},
  issue = {24},
  pages = {240601},
  numpages = {5},
  year = {2017},
  month = {Dec},
  publisher = {American Physical Society},
  doi = {10.1103/PhysRevLett.119.240601},
  url = {https://link.aps.org/doi/10.1103/PhysRevLett.119.240601}
}

@article{Bisker:17,
	doi = {10.1088/1742-5468/aa8c0d},
	url = {https://doi.org/10.1088/1742-5468/aa8c0d},
	year = 2017,
	month = {sep},
	publisher = {{IOP} Publishing},
	volume = {2017},
	number = {9},
	pages = {093210},
	author = {Gili Bisker and Matteo Polettini and Todd R Gingrich and Jordan M Horowitz},
	title = {Hierarchical bounds on entropy production inferred from partial information},
	journal = {Journal of Statistical Mechanics: Theory and Experiment}
}

@article{Jarzynski:97,
  title = {Nonequilibrium Equality for Free Energy Differences},
  author = {Jarzynski, C.},
  journal = {Phys. Rev. Lett.},
  volume = {78},
  issue = {14},
  pages = {2690--2693},
  numpages = {0},
  year = {1997},
  month = {Apr},
  publisher = {American Physical Society},
  doi = {10.1103/PhysRevLett.78.2690},
  url = {https://link.aps.org/doi/10.1103/PhysRevLett.78.2690}
}

@article{Landauer1991,
  title = {Information is Physical},
  author = {Rolf Landauer},
  journal = {Physics Today},
  volume = {44},
  issue = {5},
  pages = {23--29},
  numpages = {7},
  year = {1991},
  month = {May},
  publisher = {American Institute of Physics},
  doi = {10.1063/1.881299}
}

@article{Maes:2003,
 title={Time-Reversal and Entropy},
 author={Maes, C. and Netočný, K.},
 journal={Journal of Statistical Physics},
 volume={110},
 pages={269–310},
 year={2003},
 doi={10.1023/A:1021026930129},
 url={https://doi.org/10.1023/A:1021026930129}
}

@article{Horowitz:2009,
  title = {Illustrative example of the relationship between dissipation and relative entropy},
  author = {Horowitz, Jordan and Jarzynski, Christopher},
  journal = {Phys. Rev. E},
  volume = {79},
  issue = {2},
  pages = {021106},
  numpages = {7},
  year = {2009},
  month = {Feb},
  publisher = {American Physical Society},
  doi = {10.1103/PhysRevE.79.021106},
  url = {https://link.aps.org/doi/10.1103/PhysRevE.79.021106}
}

@article{EspositoREV,
  title = {Nonequilibrium fluctuations, fluctuation theorems, and counting statistics in quantum systems},
  author = {Esposito, Massimiliano and Harbola, Upendra and Mukamel, Shaul},
  journal = {Rev. Mod. Phys.},
  volume = {81},
  issue = {4},
  pages = {1665--1702},
  numpages = {0},
  year = {2009},
  month = {Dec},
  publisher = {American Physical Society},
  doi = {10.1103/RevModPhys.81.1665},
  url = {https://link.aps.org/doi/10.1103/RevModPhys.81.1665}
}

@article{Gaveau:2014,
  title = {Dissipation, interaction, and relative entropy},
  author = {Gaveau, B. and Granger, L. and Moreau, M. and Schulman, L. S.},
  journal = {Phys. Rev. E},
  volume = {89},
  issue = {3},
  pages = {032107},
  numpages = {5},
  year = {2014},
  month = {Mar},
  publisher = {American Physical Society},
  doi = {10.1103/PhysRevE.89.032107},
  url = {https://link.aps.org/doi/10.1103/PhysRevE.89.032107}
}

@article{JarzynskiREV,
author = {Jarzynski, Christopher},
title = {Equalities and Inequalities: Irreversibility and the Second Law of Thermodynamics at the Nanoscale},
journal = {Annual Review of Condensed Matter Physics},
volume = {2},
number = {1},
pages = {329-351},
year = {2011},
doi = {10.1146/annurev-conmatphys-062910-140506},
eprint = { https://doi.org/10.1146/annurev-conmatphys-062910-140506}
}

@article{Buscemi23,
doi = {10.1088/1367-2630/accd11},
url = {https://dx.doi.org/10.1088/1367-2630/accd11},
year = {2023},
month = {may},
publisher = {IOP Publishing},
volume = {25},
number = {5},
pages = {053002},
author = {Buscemi, Francesco and Schindler, Joseph and Šafránek, Dominik},
title = {Observational entropy, coarse-grained states, and the Petz recovery map: information-theoretic properties and bounds},
journal = {New Journal of Physics}
}

@article{Manzano24,
  title = {Thermodynamics of Computations with Absolute Irreversibility, Unidirectional Transitions, and Stochastic Computation Times},
  author = {Manzano, Gonzalo and Karde\ifmmode \mbox{\c{s}}\else \c{s}\fi{}, G\"ulce and Rold\'an, \'Edgar and Wolpert, David H.},
  journal = {Phys. Rev. X},
  volume = {14},
  issue = {2},
  pages = {021026},
  numpages = {32},
  year = {2024},
  month = {May},
  publisher = {American Physical Society},
  doi = {10.1103/PhysRevX.14.021026},
  url = {https://link.aps.org/doi/10.1103/PhysRevX.14.021026}
}

@article{Ferri25,
  title = {Conditional fluctuation theorems and entropy production for monitored quantum systems under imperfect detection},
  author = {Ferri-Cort\'es, Mar and Almanza-Marrero, Jos\'e A. and L\'opez, Rosa and Zambrini, Roberta and Manzano, Gonzalo},
  journal = {Phys. Rev. Res.},
  volume = {7},
  issue = {1},
  pages = {013077},
  numpages = {15},
  year = {2025},
  month = {Jan},
  publisher = {American Physical Society},
  doi = {10.1103/PhysRevResearch.7.013077},
  url = {https://link.aps.org/doi/10.1103/PhysRevResearch.7.013077}
}

@article{Meier25,
  title = {Emergence of a Second Law of Thermodynamics in Isolated Quantum Systems},
  author = {Meier, Florian and Rivlin, Tom and Debarba, Tiago and Xuereb, Jake and Huber, Marcus and Lock, Maximilian P.E.},
  journal = {PRX Quantum},
  volume = {6},
  issue = {1},
  pages = {010309},
  numpages = {20},
  year = {2025},
  month = {Jan},
  publisher = {American Physical Society},
  doi = {10.1103/PRXQuantum.6.010309},
  url = {https://link.aps.org/doi/10.1103/PRXQuantum.6.010309}
}

@article{Itano1993,
title = {Quantum measurements of trapped ions},
journal = {Vistas in Astronomy},
volume = {37},
pages = {169-183},
year = {1993},
issn = {0083-6656},
doi = {https://doi.org/10.1016/0083-6656(93)90029-J},
url = {https://www.sciencedirect.com/science/article/pii/008366569390029J},
author = {W.M. Itano and J.C. Bergquist and J.J. Bollinger and J.M. Gilligan and D.J. Heinzen and F.L. Moore and M.G. Raizen and D.J. Wineland}
}

@article{Keselman_2011,
doi = {10.1088/1367-2630/13/7/073027},
url = {https://doi.org/10.1088/1367-2630/13/7/073027},
year = {2011},
month = {jul},
publisher = {},
volume = {13},
number = {7},
pages = {073027},
author = {Keselman, A and Glickman, Y and Akerman, N and Kotler, S and Ozeri, R},
title = {High-fidelity state detection and tomography of a single-ion Zeeman qubit},
journal = {New Journal of Physics}
}

@article{Zhang17,
title = {Quantum feedback: Theory, experiments, and applications},
journal = {Physics Reports},
volume = {679},
pages = {1-60},
year = {2017},
issn = {0370-1573},
doi = {https://doi.org/10.1016/j.physrep.2017.02.003},
url = {https://www.sciencedirect.com/science/article/pii/S0370157317300479},
author = {Jing Zhang and Yu-xi Liu and Re-Bing Wu and Kurt Jacobs and Franco Nori}
}

@article{Lostaglio2015,
  title = {Description of quantum coherence in thermodynamic processes requires constraints beyond free energy},
  author = {Lostaglio, Matteo and Jennings, David and Rudolph, Terry},
  journal = {Nature Communications},
  volume = {6},
  issue = {1},
  pages = {6383},
  numpages = {0},
  year = {2015},
  month = {March},
  doi = {10.1038/ncomms7383},
  url = {https://doi.org/10.1038/ncomms7383}
}

@article{Scandi2020,
  title = {Quantum work statistics close to equilibrium},
  author = {Scandi, Matteo and Miller, Harry J. D. and Anders, Janet and Perarnau-Llobet, Mart\'{\i}},
  journal = {Phys. Rev. Res.},
  volume = {2},
  issue = {2},
  pages = {023377},
  numpages = {28},
  year = {2020},
  month = {Jun},
  publisher = {American Physical Society},
  doi = {10.1103/PhysRevResearch.2.023377},
  url = {https://link.aps.org/doi/10.1103/PhysRevResearch.2.023377}
}

@article{Kwon2019,
  title = {Fluctuation Theorems for a Quantum Channel},
  author = {Kwon, Hyukjoon and Kim, M. S.},
  journal = {Phys. Rev. X},
  volume = {9},
  issue = {3},
  pages = {031029},
  numpages = {26},
  year = {2019},
  month = {Aug},
  publisher = {American Physical Society},
  doi = {10.1103/PhysRevX.9.031029},
  url = {https://link.aps.org/doi/10.1103/PhysRevX.9.031029}
}

@article{Alicki_1979,
doi = {10.1088/0305-4470/12/5/007},
url = {https://doi.org/10.1088/0305-4470/12/5/007},
year = {1979},
month = {may},
publisher = {},
volume = {12},
number = {5},
pages = {L103},
author = {R. Alicki},
title = {The quantum open system as a model of the heat engine},
journal = {Journal of Physics A: Mathematical and General}
}

@article{Gong2009,
  title = {Effective Hamiltonian approach to the Kerr nonlinearity in an optomechanical system},
  author = {Gong, Z. R. and Ian, H. and Liu, Yu-xi and Sun, C. P. and Nori, Franco},
  journal = {Phys. Rev. A},
  volume = {80},
  issue = {6},
  pages = {065801},
  numpages = {4},
  year = {2009},
  month = {Dec},
  publisher = {American Physical Society},
  doi = {10.1103/PhysRevA.80.065801},
  url = {https://link.aps.org/doi/10.1103/PhysRevA.80.065801}
}

@article{Kerr1875,
author = {John Kerr},
title = {XL. A new relation between electricity and light: Dielectrified media birefringent },
journal = {The London, Edinburgh, and Dublin Philosophical Magazine and Journal of Science},
volume = {50},
number = {332},
pages = {337--348},
year = {1875},
publisher = {Taylor \& Francis},
doi = {10.1080/14786447508641302},
URL = {https://doi.org/10.1080/14786447508641302},
eprint = {https://doi.org/10.1080/14786447508641302}
}

@misc{RubinoCGQThermoCode,
  author = {Rubino, Giulia},
  title  = {{CG-QThermo}},
  year   = {2026},
  url    = {https://github.com/giulia-rubino/CG-QThermo},
  note   = {GitHub repository}
}

% Prefix a "S" to all equations, figures, tables and reset the counter
\setcounter{secnumdepth}{2}
\setcounter{section}{0}
\setcounter{equation}{0}
\setcounter{figure}{0}
\setcounter{table}{0}

% --- Appendices (REVTeX4-2 safe) ---
\appendix
\setcounter{section}{0}

% Make \ref return just A, B, C...
\renewcommand{\thesection}{\Alph{section}}

% Number equations/figures/tables as A1, A2, ... (and B1, B2, ...)
\makeatletter
\@addtoreset{equation}{section}
\@addtoreset{figure}{section}
\@addtoreset{table}{section}
\makeatother
\renewcommand{\theequation}{\thesection\arabic{equation}}
\renewcommand{\thefigure}{\thesection\arabic{figure}}
\renewcommand{\thetable}{\thesection\arabic{table}}

% Appendix section heading: "APPENDIX A. TITLE" (TITLE in caps)
\newcommand{\appsection}[1]{%
  \refstepcounter{section}%
  \section*{APPENDIX \thesection.\ \MakeUppercase{#1}}%
}

\appsection{Coarse-graining and information thermodynamics}
\label{sec:thermocost}

In modern approaches to thermodynamics, the second law is considered as a consequence of the fundamental link between energetics and information theory~\cite{Parrondo2015,Goold_2016}. One of the most prominent examples of this strong connection was noticed by R. Landauer. Under the popular lemma `information is physical', Landauer developed the idea that logically irreversible operations, such as the erasure of a memory, have an associated thermodynamic cost~\cite{Landauer1991}, hence providing a route for exorcising Maxwell's demon paradox~\cite{Mayurama:09}. Here, we explore the thermodynamic implications of having different levels of information as determined by the access to coarse-grained or fine-grained instruments with different levels of energy resolution.

\subsection{Extracting work using fine-grained information}

Let us now consider two agents, $C$ and $F$, equipped with instruments of differing energy resolution. The coarse-grained agent $C$, whose instruments have limited precision, models the system using the coarse-grained Hamiltonian $\breve{H}$ defined in Eq.~\eqref{eq:coarseH}. When the system is brought into contact with a thermal reservoir at inverse temperature $\beta$, $C$ describes its equilibrium state as $\breve{\tau}_\beta$ [Eq.~\eqref{eqn:expRho}]. In contrast, the fine-grained agent $F$ has access to higher-resolution instruments and assigns to the same system the fine-grained Hamiltonian $H = \sum_j \epsilon_j \ket{\epsilon_j}\bra{\epsilon_j}$. Accordingly, from $F$'s perspective, the equilibrium state of the system is $\tau_\beta$ [Eq.~\eqref{eqn:thermal}]~\footnote{Here we are implicitly assuming that the interaction with the reservoir is such that it is able to thermalise all degrees of freedom, including the fine-grained ones. This assumption is usually referred to as ``ergodicity" of the system.}.
The origin of this discrepancy lies in the fact that agent $C$ cannot distinguish $\tau_\beta$ from $\breve{\tau}_\beta$: from their perspective, the two states are operationally equivalent. However, from agent $F$’s viewpoint, the states are markedly different. In particular, $F$ regards $\breve{\tau}_\beta$ as a non-equilibrium state, and thus as a potential thermodynamic resource~\cite{Goold_2016}. 

Let us now consider the situation in which the state $\breve{\tau}_\beta$ is \emph{precisely} prepared by some means and handed to agent $F$. With access to high-resolution instruments, $F$ may then extract work from $\breve{\tau}_\beta$ by implementing a thermodynamic protocol that cyclically manipulates the system’s Hamiltonian $H$, thereby transforming $\breve{\tau}_\beta \rightarrow \tau_\beta$ while the system remains in contact with the environment~\cite{Parrondo2015}. The maximum amount of work extractable in such a process is bounded by the free energy difference between the initial and final states~\cite{Parrondo2015,Anders2016,Popescu2014}:
\begin{equation}\label{eq:workext}
W_{\rm ext} \leq \mathcal{F}(\breve{\tau}_\beta) - \mathcal{F}(\tau_\beta) = k_B T \, S(\breve{\tau}_\beta || \tau_\beta),
\end{equation}
where $\mathcal{F}(\rho) \equiv \tr[H \rho] - k_B T S(\rho)$ is the non-equilibrium free energy of a generic state $\rho$ with respect to the Hamiltonian $H$, which reduces to the equilibrium free energy $F$ for thermal states, and we have $\mathcal{F}(\rho) \geq \mathcal{F}(\tau_\beta)= F$ for any non-equilibrium state $\rho$. Above, in the r.~h.~s., we introduced the quantum relative entropy (or Kullback-Leibler divergence), $S(\rho||\sigma) \equiv \tr[\rho (\log \rho - \log \sigma)]$, a non-negative  information-theoretic measure of the distinguishability between quantum states~\footnote{More precisely, given $n$ copies of a state $\sigma$, the probability of guessing incorrectly that the data derive from $\rho$ is bounded by $2^{-n S(\rho||\sigma)}$ for large $n$~\cite{VedralRev}.}.

The bound in Eq.~\eqref{eq:workext} can be saturated (i.e., maximum work can be extracted) in the reversible limit (zero entropy production)~\cite{Parrondo2015}. 
The latter can be achieved, for example, by performing a sudden quench $H \rightarrow H' \equiv -k_B T \log \breve{\tau}_\beta$, followed by an isothermal quasi-static modulation of the Hamiltonian back to its initial form, $H' \rightarrow H$, while remaining in equilibrium with the thermal environment~\cite{Takara:10,Esposito2011}. We remark that such work extraction protocol requires control over the fine-grained energy levels $\ket{\epsilon_i}$ of $H$ within the slots $\mathcal{J}_J$, and therefore can only be done by using fine-grained instruments. Eq.~\eqref{eq:workext} highlights the fact that a greater ability to detect and manipulate fine-grained degrees of freedom comes with an increased ability to extract work, in the same spirit as in Maxwell's demon. 

Now consider the reverse situation, in which the fine-grained thermal state $\tau_\beta$ is prepared and handed to the coarse-grained agent $C$. Suppose that $C$, attempting to replicate the protocol used by agent $F$ in the previous scenario, aims to extract work from $\tau_\beta$ using their coarse-grained instruments. Due to the limited resolution of these instruments, however, agent $C$ can only access the state after the coarse-graining operation has been applied, namely, $\mathcal{C}_{\Pi_J}(\tau_\beta)$. Applying an analogous protocol to the one described above, agent $C$ would then extract an amount of work given by:
\begin{equation}\label{eq:workCG}
\breve{W}_\mathrm{ext} = k_B T \, S\bigl(\mathcal{C}_{\Pi_J}(\tau_\beta) || \breve{\tau}_\beta\bigr) = 0,    
\end{equation}
where the last equality is verified whenever $\mathcal{C}_{\Pi_J}(\tau_\beta) = \breve{\tau}_\beta$, as required in the previous section, c.f.~Eq.~\eqref{eq:condition}. It is worth remarking the importance of such a condition to ensure thermodynamic consistency: if $\mathcal{C}_{\Pi_J}(\tau_\beta) \neq \breve{\tau}_\beta$, the work extractable by the coarse-grain agent $C$ in Eq.~\eqref{eq:workCG} would be strictly positive. If that was possible, then a cycle could be constructed where agent $F$ first prepares $\tau_\beta$ by thermalising the system in contact with the thermal bath and then agent $C$ extracts work from it, giving the system back to $F$ for a new thermalisation. Repeating such a cycle would lead to extracting an infinite amount of work from a single thermal bath, hence violating the second law of thermodynamics.

\subsection{Free energy loss from coarse-graining and work cost}

In the above, we have seen that access to fine-grained degrees of freedom can increase one's ability to extract work. 
In particular, we have seen that a positive amount of work can be extracted by agent $F$ by transforming $\breve{\tau}_\beta$ into $\tau_\beta$. We now turn the situation around and focus on the opposite transformation, $\tau_\beta \rightarrow \breve{\tau}_\beta$, that is, the preparation of state $\breve{\tau}_\beta$. Interestingly, such an operation is quite different from the perspective of agent $F$ or agent $C$. 
For example, if agent $F$ prepares state $\tau_\beta$ and gives it to agent $C$, the transformation $\tau_\beta \rightarrow \mathcal{C}_{\Pi_J}(\tau_\beta) = \breve{\tau}_\beta$ occurs spontaneously from the loss of precision entailed by the coarse-grained instruments, leading to a drop in the free energy from agent $F$ perspective:
\begin{equation}
    \breve{\mathcal{F}}(\breve{\tau}_\beta) - \breve{\mathcal{F}}(\tau_\beta) = S({\tau}_\beta) - S(\breve{\tau}_\beta) \leq 0,
\end{equation}
where we denoted $\breve{\mathcal{F}}(\rho)$ the free energy with respect to $\breve{H}$. The equality follows from energy conservation, $\tr[\breve{H} \breve{\tau}_\beta] = \tr[\breve{H} \mathcal{C}_{\Pi_J}({\tau}_\beta)] = \tr[\breve{H} {\tau}_\beta]$, and the inequality from the fact that unital channels are entropy non-decreasing, hence $S(\breve{\tau}_\beta) = S(\mathcal{C}_{\Pi_J}(\tau_\beta)) \geq S({\tau}_\beta)$.

In contrast, a \emph{precise} preparation of state $\breve{\tau}_\beta$ from agent $F$’s perspective is a logically irreversible operation with thermodynamic consequences and an associated work cost, analogous to what happens in information erasure scenarios~\cite{Mayurama:09,Parrondo2015}. In particular, if the system is put in contact with a thermal bath at inverse temperature $\beta$, agent $F$ will assign state $\tau_\beta$ to it, not $\breve{\tau}_\beta$. To prepare $\breve{\tau}_\beta$ instead, $F$ must apply additional manipulations to transform $\tau_\beta$ into the desired state. Assuming control over the Hamiltonian $H$ and access to the bath, that transformation can be done in an optimal way by combining again sudden quenches of $H$ with quasi-static modulations (see Appendix~\ref{app:B} for details), which leads to a positive work cost lower-bounded by:
\begin{align}\label{eq:workcost}
    W_{\rm cost} &\geq \mathcal{F}(\breve{\tau}_\beta) - \mathcal{F}(\tau_\beta) = k_B T S(\breve{\tau}_\beta || \tau_\beta).
\end{align}
Notice that in opposition to agent $C$, from agent $F$ perspective, the coarse-grained state $\breve{\tau}_\beta$ has greater free energy than $\tau_\beta$, which follows from the non-negativity of the quantum relative entropy $S(\breve{\tau}_\beta || \tau_\beta) \geq 0$. This is indeed the reason why $F$ is able to extract work from $\breve{\tau}_\beta$ as captured by Eq.~\eqref{eq:workext} above. However, now we see that preparing the state $\breve{\tau}_\beta$ \emph{precisely} using fine-grained instruments has an equivalent work cost.

The r.~h.~s.~of Eq.~\eqref{eq:workcost} provides the minimal work cost for preparing $\breve{\tau}_\beta$ in a fine-grained perspective from the true thermal state $\tau_\beta$, $W_{\rm min} := k_B T S(\breve{\tau}_\beta || \tau_\beta)$. An upper bound to that minimal cost can be obtained by using again the unitality of the map $\mathcal{C}_{\Pi_J}$, which implies that the change in non-equilibrium free energy is upper bounded by the average energy change under coarse-graining:
\begin{equation} \label{eq:upbound}
W_{\min} \leq \tr[H \, \breve{\tau}_\beta] - \tr[H \tau_\beta] \leq \delta \epsilon.  
\end{equation}
The last inequality above is proven in Appendix~\ref{app:B} and shows that the (minimal) work cost remains below the precision of the coarse-grained instruments, $\delta \epsilon$, and therefore cannot be detected by agent $C$. Nevertheless, we stress that non-optimal irreversible preparations of $\breve{\tau}_\beta$ may lead to higher work costs that might also be detectable with coarse-grained instruments.

Finally, we emphasize that the work cost in Eq.~\eqref{eq:workcost} needs to be paid if the state of the system is transformed from the point of view of agent $F$. Another way to view why there should be a cost in that situation is to consider that the transformation would require forgetting the information about the fine-grained degrees of freedom. This is in contrast to the viewpoint of agent $C$, which would ascribe the same (macro)state $\breve{\tau}_\beta$ to many different fine-grained states, including $\tau_\beta$, and therefore would not be able to detect any physical transition $\tau_\beta \rightarrow \breve{\tau}_\beta$.
%any \emph{apparent} transformation, occurring from the point of view of $C$, due to an increase in the ignorance of the fine-grained degrees of freedom. 
However, such ignorance of the system's state may lead to \emph{real} consequences when manipulating the system, as we shall see in the following sections.

%%%%%%%%%%%%%%%%%%%%%%%%%
%\clearpage
%%%%%%%%%%%%%%%%%%%%%%%%%

%\onecolumngrid
%\begin{center}
%\textbf{\large Supplementary Material}
%\end{center}
%
%\vskip 0.2 cm
%
%\onecolumngrid
%
% Prefix a "S" to all equations, figures, tables and reset the counter
%\setcounter{secnumdepth}{2}
%\setcounter{section}{0}
%\setcounter{equation}{0}
%\setcounter{figure}{0}
%\setcounter{table}{0}
%%\setcounter{page}{1}
%\makeatletter
%\renewcommand{\theequation}{S\arabic{equation}}
%\renewcommand{\thefigure}{S\arabic{figure}}
%\renewcommand{\thesection}{S.\Roman{section}}
%%\renewcommand{\bibnumfmt}[1]{[S#1]}
%%\renewcommand{\citenumfont}[1]{S#1}

\appsection{Work cost for the fine-grained preparation of coarse-grained states}
\label{app:B}
In this Appendix, we detail the construction of an optimal thermodynamic protocol for preparing the coarse-grained thermal state $\breve{\rho}_\beta$ from an arbitrary initial state $\rho$, using fine-grained instruments.

The protocol follows the approach used in Refs.~\cite{Takara:10,Anders:13,Manzano2018}, adapted to the present setting. It consists of the following three steps: \textit{i.}~A first sudden quench of the Hamiltonian, $H \rightarrow H_0 \equiv -k_B T\log \rho$, which induces no change in the system state $\rho$. \textit{ii.}~A quasi-static modulation of the Hamiltonian from $H_0 \rightarrow H_1 \equiv - k_B T \log \mathcal{C}(\rho)$, while the system is maintained in thermal contact with the environment. \textit{iii.}~A second sudden quench transforming back $H_1 \rightarrow H$ without affecting the system state. In the above sequence, steps~\textit{i.}~and~\textit{iii.}~occur on a timescale much shorter than the system's typical relaxation time into the reservoir, while step~\textit{ii.}~is assumed to be a quasi-static evolution, slow compared to the relaxation timescale. Consequently, the action of the environment can be neglected during steps~\textit{i.}~and~\textit{iii.}, and the system state remains unchanged, resulting in zero entropy change $\Delta S_1 = \Delta S_3 = 0$. The energy changes in these sudden quenches can thus be interpreted as work, $W_1 = \tr[(H_0 - H) \rho]$, and $W_3 = \tr[(H - H_1) \, \mathcal{C}(\rho)]$. In step~\textit{ii}, by contrast, both heat and work are exchanged, with the second law imposing that $Q_2 \leq k_B T \Delta S_2$, where $\Delta S_2$ is the system's entropy change during the quasi-static evolution. When the modulation is sufficiently slow, the system remains approximately in thermal equilibrium at each moment, following a sequence of thermal states of the instantaneous Hamiltonian at inverse temperature $\beta$. In this quasi-static limit, the process becomes reversible, and the heat exchanged with the environment equals the entropy change times the temperature, $Q_2 = k_B T [S(\mathcal{C}(\rho)) - S(\rho)]$, where we used the fact that $\mathcal{C}(\rho) = e^{-\beta H_1}$. The work performed in this step is then $W_2 = \Delta E_2 - Q_2$, with $\Delta E_2 = \tr[H_1 \mathcal{C}(\rho)] -\tr[H_0 \rho)]$. Summing up the three contributions, the total work invested in the protocol reads:
\begin{align}\label{eq:workcost_app}
    W_{\rm cost} \geq  W_{\min} &= F(\mathcal{C}(\rho)) - F(\rho)
    \notag\\
    &= k_B T \bigl[S(\mathcal{C}(\rho) || \tau_\beta) - S(\rho || \tau_\beta)\bigr],
\end{align}
where the inequality in the first line follows from the fact that, in general, finite-time protocols absorb heat $Q_2 \leq \Delta S_2$, and hence $W_2 \geq \Delta E_2 - k_B T \Delta S_2$ (see also below). In the second line of Eq.~\eqref{eq:workcost_app}, we used the fact that $\tr[H \sigma] = -k_B T \tr[\sigma \log \tau_\beta] - k_B T \log Z$ for $\sigma = \rho, \mathcal{C}_{\Pi_J}(\rho)$.

Eq.~\eqref{eq:workcost} implies that, in order to prepare $\breve{\tau}_\beta$ at the fine-grained level, a non-zero input work is needed whenever the state $\mathcal{C}_{\Pi_J}(\rho)$ is farther from equilibrium than the initial state $\rho$. This difference is accounted for by the distinguishabilities of $\mathcal{C}_{\Pi_J}(\rho)$ and $\rho$ from the thermal state $\tau_\beta$ through the quantum relative entropy. In particular, when the initial state is already thermal, i.e., $\rho = \tau_\beta$, the second term on the right-hand side of Eq.~\eqref{eq:workcost} vanishes, and we recover the (strictly positive) cost $W_{\rm cost} \geq k_B T \, S(\breve{\tau}_\beta|| \tau_\beta)$ as reported in Eq.~\eqref{eq:workcost}, which matches the extractable work in the reverse process (cf.~Eq.~\eqref{eq:workext}).

The minimal work cost in Eq.~\eqref{eq:workcost_app} is achieved when the protocol described above is implemented with a clear separation of timescales between the system's relaxation into the thermal bath and the modulation of the Hamiltonian. This corresponds to a reversible process, characterised by zero entropy production. In all other cases, the coarse-graining operation results in nonzero entropy production:
\begin{align}
\Sigma &= S[\mathcal{C}_{\Pi_J}(\rho)] - S(\rho)  - \beta Q  \notag \\
&= \beta \Big( W_{\rm cost} - F[\mathcal{C}_{\rm cost}(\rho)] + F(\rho) \Big) \geq 0,    
\end{align}
where $Q$ is the total heat absorbed by the system during the process. This additional entropy production comes with an extra work cost, leading to the inequality $W_{\rm cost} \geq W_{\min}$ in Eq.~\eqref{eq:workcost_app}.

Finally, we prove that the energy change in the system due to the coarse-graining operation, $\Delta E_\mathrm{C} := \tr[H C_{\Pi_J}(\rho)] - \tr[H \rho]$, lies below the minimum energy resolution of their experimental instruments:
\noindent\parbox{\columnwidth}{
\begin{align}
    \Delta E_\mathcal{C} &= \sum_J \big(\bar{E}_J \tr[\rho \Pi_J] - \sum_{j \in \mathcal{J}_J} \epsilon_j \langle \epsilon_j | \rho | \epsilon_j \rangle \big) \\
    &\leq \sum_J (\bar{E}_J - E_J^{\min})\tr[\rho \Pi_J] \leq \delta \epsilon \sum_J \tr[\rho \Pi_J] = \delta \epsilon. \notag
\end{align}}
The above inequality implies that, from the perspective of agent $C$, the associated energy change is negligible. Consequently, by the left-hand inequality in Eq.~\eqref{eq:upbound}, ${W}_{\min} \leq \Delta E_\mathcal{C}$, the minimum work cost of coarse-graining is also negligible for agent $C$.

\appsection{Effect of commuting driving on coarse-grained work statistics}
\label{app:C}

One might wonder whether different behaviour would arise if the driving and initial Hamiltonians commuted at all times, meaning that no transitions between energy eigenstates would be induced by the dynamics. In this case, the evolution would preserve the energy eigenbasis, and the dynamics could be considered effectively classical as populations would not be redistributed by the unitary evolution.

To explore this scenario, we examine a modified driving protocol where the driven Hamiltonian is expressed as $H_\tau = \hbar\omega(a^\dagger a + 1/2) + \hbar \chi N^2$, where $N = a^\dagger a$ and $\chi$ is the third-order susceptibility~\cite{Gong2009}.
This term is formally analogous to the Kerr nonlinearity found in quantum optics, where an intensity-dependent refractive index gives rise to an energy shift proportional to the square of the photon number~\cite{Kerr1875}. We therefore refer to this as a Kerr-type commuting drive.
Since $[H_0, H_\tau] = 0$, the time evolution operator is diagonal in the energy eigenbasis and transitions between different energy levels do not occur.

The resulting coarse-grained dissipative work, shown in Fig.~\ref{img:DissWork_RelEntropy_Classical}, exhibits trends that are highly similar to those obtained for the non-commuting drive discussed in Sec.~\ref{subsec:Case_Study}. In particular, the dependence on driving strength and coarse-graining resolution follows the same qualitative behaviour. This suggests that the main features observed in the coarse-grained thermodynamic quantities are not primarily influenced by the occurrence of quantum transitions between energy eigenstates.

\begin{figure}[bt]
\centering
\includegraphics[width=\columnwidth]{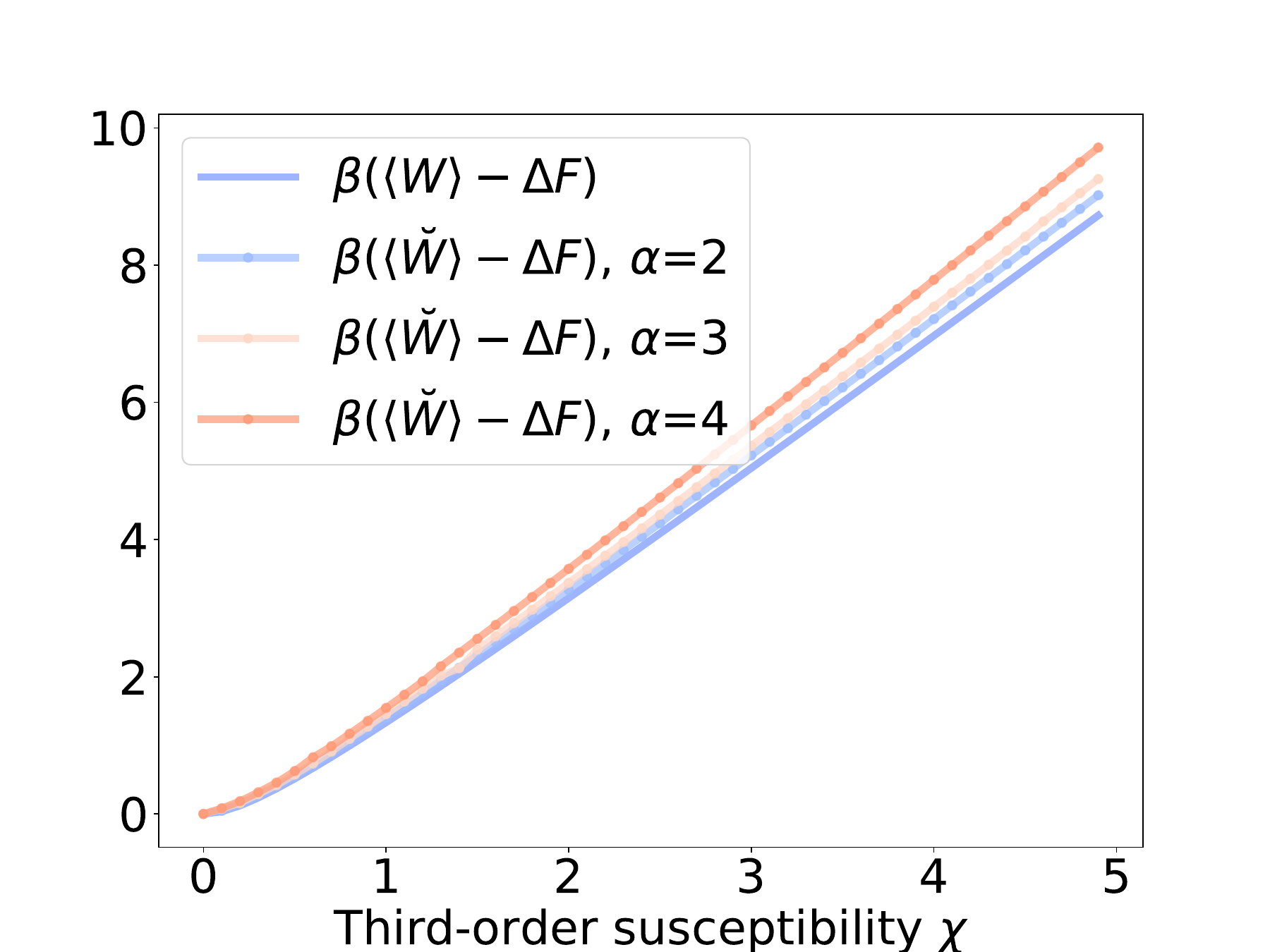}
\captionof{figure}{\footnotesize{Trends of fine- and coarse-grained dissipative work for a harmonic oscillator under a Kerr-type commuting drive.}
The initial state is the thermal state of the harmonic oscillator Hamiltonian $H_0 = \hbar\omega(a^\dagger a + 1/2)$, and the final state is generated by time-independent, number-conserving driving applied for $t\in(0,\tau]$, such that $H_\tau = \hbar\omega(a^\dagger a + 1/2) + \hbar\chi N^2$, satisfying $[H_0,H_\tau] = 0$. The quantities plotted and all parameters are the same as in Fig.~\ref{img:DissWork_RelEntropy}, with coarse-graining performed analogously using the same resolution parameter $\alpha$. Despite the absence of quantum transitions between energy eigenstates, the qualitative dependence of dissipative work on $\chi$ and on the coarse-graining resolution closely mirrors that of the non-commuting driving case. This indicates that the observed coarse-graining effects are largely insensitive to whether the driving induces transitions or only reshapes the energy spectrum.}
\label{img:DissWork_RelEntropy_Classical}
\end{figure}

This similarity can be explained by the fact that the effects reported in the main text are primarily governed by the coarse-graining of the energy spectrum and the thermal weighting within each coarse-grained energy slot rather than by the presence or absence of quantum transitions between fine-grained energy eigenstates. Consequently, coarse-grained thermodynamic quantities, such as average work and free-energy differences, are largely insensitive to whether the underlying dynamics generate coherences, provided the same coarse-graining prescription is used.
Nevertheless, the observed similarity in coarse-grained thermodynamic quantities does not imply that the underlying dynamics are equivalent. In particular, quantities that probe the structure of the transition probabilities, such as the presence of off-diagonal contributions or measures of transition-induced mixing, vanish identically in the commuting case, yet are generally non-zero when the driving does not commute with the Hamiltonian. While these differences are washed out by coarse-graining at the level of average thermodynamic quantities, they highlight the genuinely quantum nature of non-commuting dynamics and clarify in what sense coarse-graining can obscure distinctions between classical and quantum driving scenarios.

\appsection{Coarse graining as limited energy resolution}
\label{app:D}

In this work, the coarse-graining procedure is restricted to being diagonal in the energy eigenbasis. This is not just a technical simplification, but a physical modelling assumption tailored to the thermodynamic scenario of interest. The framework is designed to capture how thermodynamic quantities and relations are modified when the system's energy can only be resolved up to finite experimental precision, rather than to describe arbitrary forms of information loss on the quantum state.

More generally, one could also consider coarse-graining procedures acting on other microscopic degrees of freedom, implemented through projective measurements that are not functions of the Hamiltonian. This type of coarse-graining can be described using the same mathematical framework adopted here: projectors can be used to define a partition of Hilbert space, along with an associated coarse-graining map. However, whether thermodynamic relations remain valid in this case depends crucially on the commutation properties between the coarse-grained observables and the Hamiltonian. When coarse-graining does not commute with energy, it generally mixes energy populations and coherences, introducing additional physical disturbance rather than merely reflecting limited measurement resolution. This type of mixing is typical of dynamics generated by interactions that do not commute with the Hamiltonian and has observable consequences for work statistics and entropy production~\cite{Lostaglio2015,Scandi2020,Kwon2019}. As a result, such coarse graining requires a separate thermodynamic analysis and cannot be interpreted simply as a restriction on the observer’s access to information.

By contrast, one may consider more general coarse-graining schemes that still probe the energy but with finite accuracy, rather than restricting to sharp projective measurements. A first natural extension is to allow for unsharp measurements that remain diagonal in the energy basis, described by POVMs with overlapping effects, which model readout noise or imperfect calibration. As discussed in Sec.~\ref{sec:GeneralizationToPOVM}, the associated coarse-graining map is then no longer idempotent and coarse-grained thermal states do not generally retain an exact Gibbs form, so several of the exact thermodynamic identities derived here no longer hold without further assumptions.

Another class of generalisations consists of channels that preserve energy populations, but mix coherences within degenerate subspaces or between nearby energies. This may arise from uncontrolled, yet energy-preserving, interactions with an environment. Although these processes leave the average energy unchanged, they modify work statistics and generally alter fluctuation relations. This requires a reformulation of the thermodynamic framework, as evidenced by studies showing that quantum coherence contributes additional terms to work distributions~\cite{Aberg2018,Scandi2020} and affects fluctuation theorems for general quantum channels~\cite{Manzano18b,Kwon2019}.

Finally, fully general coarse-graining operations that do not respect the energy eigenspaces correspond to open-system dynamics rather than to limited measurement resolution. In this regime, neither thermal fixed points nor work definitions based on the system Hamiltonian are preserved~\cite{EspositoREV,Alicki_1979}, and the resulting entropy production cannot be attributed solely to coarse graining. Investigating thermodynamic relations under such general dynamical coarse graining is therefore a distinct problem, beyond the scope of the present work.

\end{document}